\documentclass[aps,prd,reprint,twocolumn,nofootinbib,superscriptaddress,preprintnumbers,floatfix
]{revtex4-1}
\usepackage{amsmath}
\usepackage{amsfonts}
\usepackage{amssymb}
\usepackage{graphicx}
\usepackage{epsfig}
\usepackage{subfigure}
\usepackage{color}
\usepackage{float}

\newcommand{\be}{\begin{equation}}
\newcommand{\ee}{\end{equation}}
\newcommand{\ba}{\begin{eqnarray}}
\newcommand{\ea}{\end{eqnarray}}

\def\Li{\textrm{Li}}
\def\ln{\textrm{ln}}

\def\nn{\nonumber}
\def\MS{\overline{\rm MS}}

\begin{document}


\preprint{\vbox{\hbox{UWTHPH 2013-16}
}}

\title{\Large Two loop soft function for secondary massive quarks}

\author{Simon Gritschacher}
\affiliation{Mathematisches Institut, Ludwig-Maximilians-Universit\"at M\"unchen,
Theresienstra\ss{}e 39, 80333 M\"unchen, Germany}

\author{Andre H. Hoang, Ilaria Jemos and Piotr Pietrulewicz}
\affiliation{Fakult\"at f\"ur Physik, Universit\"at Wien,
Boltzmanngasse 5, 1090 Vienna, Austria}

\begin{abstract}
We present the calculation of the $\mathcal{O}(\alpha_s^2 C_F T_F)$ massive quark corrections to the soft function for the double hemisphere jet mass distribution in $e^+ e^-$ collisions, a necessary ingredient for the calculation of several event shape distributions at N${}^3$LL order. The use of the mass as an infrared regulator allows us to derive the momentum space results for the massless quark structures at $\mathcal{O}(\alpha_s^2 C_F T_F n_f)$ and the gluonic structures at $\mathcal{O}(\alpha_s^2 C_A C_F)$, which have not been given so far in the literature. Furthermore, we compute the corresponding corrections in the soft function for thrust, the most prominent projection of the double hemisphere mass distribution. Finally we give expressions for the corresponding renormalon subtractions in the gap scheme.
\end{abstract}
\maketitle

\section{Introduction}

The theoretical description of event shape distributions in $e^+e^-$ collisions
has recently seen substantial progress concerning the treatment of higher-order
QCD corrections~\cite{Weinzierl:2008iv,Weinzierl:2009ms,GehrmannDeRidder:2007bj,GehrmannDeRidder:2007hr}, the techniques concerning the summation of large
logarithmic terms~\cite{Becher:2008cf,Chien:2010kc,Becher:2012qc,Chiu:2011qc,Chiu:2012ir,Gehrmann:2012sc} and the implementation of schemes that avoid
renormalon ambiguities together with the definition of non-perturbative
parameters~\cite{Hoang:2007vb,Hoang:2009yr}. These developments have contributed to an improved
theoretical accuracy for the description of event shape distributions and to
precise measurements of QCD parameters such as the strong
coupling~\cite{Abbate:2010xh,Abbate:2012jh,Gehrmann:2009eh,Gehrmann:2012sc}.  An important development for making higher order
corrections accessible in a systematic way is the framework of Soft-Collinear
Effective Theory (SCET)~\cite{Bauer:2000ew,Bauer:2000yr} which makes it possible to factorize the most singular
contributions for a large class of event shapes in the dijet limit in terms of a
hard current Wilson coefficient, a jet function describing the collinear
radiation and a soft function describing large-angle soft radiation. Within the
SCET framework it has also become possible to treat coherently the effects of
the production of massive quarks which are the focus of this work.  
In $e^+e^-$-collisions one can distinguish two types of heavy quark production mechanisms: {\it primary}, where the heavy quarks are produced directly by the hard current, and {\it secondary}, where massless quarks are
produced by the hard current and the heavy quarks arise from gluon splitting.

For the production of primary heavy quarks, factorization in the dijet limit for the
c.m.\ energy being much larger than the mass was discussed  in
Ref.~\cite{Fleming:2007qr} and results suitable for a description at NNLL
order\footnote{  
In this work we use standard SCET counting where NNLL order refers to the cusp
anomalous dimension at ${\cal O}(\alpha_s^3)$, the non-cusp anomalous dimension
at ${\cal O}(\alpha_s^2)$ and fixed-order matrix element corrections at ${\cal
  O}(\alpha_s)$.}~\cite{Fleming:2007xt}  
were provided for the case that the quark mass is of order of the jet invariant mass. An important conceptual finding of Ref.~\cite{Fleming:2007xt}  
was that, as long as only secondary radiation of massless partons is considered,
the soft function remains unchanged with respect to the case of primary 
massless quark production in the dijet limit, so that only the jet function
receives non-trivial modifications due to the heavy quark mass. It was further
shown for the same situation that approaching the heavy quark production
threshold in the collinear sector when the off-shellness is much smaller than
the heavy quark mass, an additional matching onto boosted versions of
Heavy-Quark Effective Theory (HQET) is required, which does not affect the soft sector.  

For the production of secondary heavy quarks no
coherent approach of how the quark mass affects factorization has been presented
until recently. While it was known that at LL and NLL order the main effect is
related to the number of active running quark flavors in the evolution equations
for the strong coupling and the renormalization group factors in the
factorization theorem~\cite{Fleming:2007qr,Kaplan:2008pt}, the conceptual background of how to go beyond NLL
order, which includes non-trivial matrix element corrections and matching conditions was only
provided recently in Ref.~\cite{Gritschacher:2013pha}. In that work it was shown that the problem
of secondary heavy quark production is closely related to the problem of massive
gauge boson production in jet observables, because the production of a heavy
quark-antiquark pair off a virtual gluon can be calculated from a dispersion
integral over the gluon invariant mass. In addition to the usual collinear and
soft degrees of freedom known for the purely massless case, the resulting
factorization framework requires so-called {\it mass modes}, which are collinear
and soft degrees of freedom with a common typical invariant mass of the order of the
heavy quark mass. The mass modes are integrated out when the evolution crosses
the heavy quark mass threshold and allow for a continuous description of the
singular terms from infinitely heavy down to infinitesimally small masses
merging into the known massless limit. It was also demonstrated in
Ref.~\cite{Gritschacher:2013pha} that when the mass modes are integrated out the
associated matching conditions in the collinear and soft sectors can involve
non-trivial plus-distributions in the respective kinematic variables.

The above mentioned dispersion method is exact for cases where the momenta of
the produced massive quark and antiquark momenta enter the observable in a
coherent manner such as for the calculation of the jet function or for purely
virtual corrections like the ones entering the hard current Wilson
coefficient. On the other hand, for the soft function, where the two quarks can
enter into different hemispheres and their momenta contribute incoherently due
to phase space constraints, the dispersion method does not lead to the correct
finite non-logarithmic corrections. Thus the ${\cal O}(\alpha_s^2 C_F T_F)$
massive quark corrections to the soft function have to be determined by a
dedicated computation along the lines of Ref.~\cite{Kelley:2011ng} where the ${\cal
  O}(\alpha_s^2)$ corrections to the soft function from gluons and massless
quarks were determined by lengthy calculations (see also Ref.~\cite{Hornig:2011iu} for a discussion of non-global
logarithms at ${\cal O}(\alpha_s^2)$ as well as~\cite{Hoang:2008fs}). It is the main aim of this paper to
present and discuss the  ${\cal O}(\alpha_s^2 C_F T_F)$ massive quark
corrections to the double hemisphere soft function and to outline their
calculation. The results are important for event shapes such as thrust and the
heavy jet mass. We assume for the most part of this paper that the heavy quark
mass is of the order of the 
typical soft momenta, so we consider virtual as well as real corrections due to
secondary massive quark production accounting for the exact analytic threshold
behavior.  

An interesting conceptual feature of the result is that the quark mass serves as a
physical infrared regulator which provides a manifest separation of IR-sensitive
and UV-divergent structures. This separation is less obvious and more difficult
to make manifest for the massless case when IR and UV divergences are
regularized by dimensional regularization. Using the ${\cal O}(\alpha_s^2 C_F
T_F)$ massive quark corrections to the soft function and the fact that the
UV-divergences agree with the massless quark case, it is possible to determine
the distributive analytical structure of the  ${\cal O}(\alpha_s^2 C_F T_F n_f)$
massless quark corrections to the momentum space double hemisphere soft function. Taking these steps as a guideline one can then also deduce the momentum space representation for all ${\cal O}(\alpha_s^2)$ corrections. This
analytical distributive structure of the momentum space double hemisphere soft function was
not identified in Refs.~\cite{Kelley:2011ng,Hornig:2011iu} and represents 
an additional result of this work.  

The finite quark mass also provides a physical cutoff against infrared
renormalons that arise for massless quarks in high order corrections and
enhance the sensitivity to small gluon virtuality. In this work we nevertheless
discuss the subtraction of the perturbative ${\cal O}(\Lambda_{\rm QCD})$
renormalon contributions along the lines of Refs.~\cite{Hoang:2007vb,Jain:2008gb,Hoang:2008fs} for the $\mathcal{O}(\alpha_s C_F T_F)$ massive quark corrections. The knowledge
of this subtraction is required in cases when the quark mass decreases below the scale
of the soft radiation in order to achieve a continuous transition to the massless
approximation which has been used in many previous analyses. It is also required for the determination of the matching
condition when the R-evolution in the renormalon-subtracted scheme~\cite{Hoang:2008yj,Hoang:2009yr} for
the soft power correction~\cite{Hoang:2007vb,Jain:2008gb,Hoang:2008fs} crosses the mass threshold. 

The content of this paper is as follows: In Sec.~\ref{sec:factorization} we summarize briefly the SCET factorization theorem for double hemisphere masses in the dijet region. Since the computation of the soft hemisphere function involves several steps we first give an outline of the method we use in Sec.~\ref{sec:outline} followed by the explicit details of the corresponding phase space calculations given in Secs.~\ref{sec:gluonhemisphere} and~\ref{sec:quarkhemisphere}. In Sec.~\ref{sec:massless} we use our results with massive quarks to derive explicit expressions for the massless limit of the soft hemisphere function. As an explicit example for an event shape derived from the hemisphere masses we discuss the massive thrust soft function in Sec.~\ref{sec:projections}. In Sec.~\ref{sec:renormalon_subtraction} we present the results for the corresponding renormalon subtractions of the massive soft function, before we conclude in Sec.~\ref{sec:conclusions}.

\section{Factorization Theorem and Soft Function Definition}\label{sec:factorization}

We focus on the dijet invariant mass distribution in $e^+ e^-$ annihilation, where one defines two hemispheres
(left/right) separated by the plane perpendicular to the thrust axis
$\mathbf{n}$. 
The factorization theorem for the singular terms in the dijet limit, where the c.m.energy $Q$ is much larger than the jet masses, accounting only for massless secondary quarks has the form~\cite{Fleming:2007qr,Fleming:2007xt}
\begin{align}\label{eq:doublehemifactorization}
 \frac{1}{\sigma_0}\frac{d^2 \sigma}{dM_l^2 dM_r^2}=& \,H(Q,\mu^2)\int dk_l \,dk_r
 \,J_n(M_l^2-Qk_l,\mu) \nn \\
 & \times \,J_{\bar{n}}(M_r^2-Qk_r,\mu)\, S(k_l,k_r,\mu) \, , 
\end{align}
where the hemisphere mass $M_l$ ($M_r$) denotes the invariant mass in the
left (right) hemisphere. The terms $H(Q,\mu)$, $J_{n,\bar{n}}(s,\mu)$,
$S(k_l,k_r,\mu)$ are the hard, jet and soft functions and all renormalization group
factors which sum large logarithms are implicit. 
The knowledge of the double differential distribution allows us to derive the
differential cross sections for several other event shape variables like thrust
via 
\begin{align}\label{eq:thrust_dist}
\frac{d\sigma}{d\tau}=\int dM_l^2 dM_r^2 \,\frac{d^2 \sigma}{dM_l^2 dM_r^2} \,\delta\left(\tau-\frac{M_l^2}{Q^2}-\frac{M_r^2}{Q^2}\right)
\end{align}
in the dijet limit. 

In the case of massive secondary particles the basic structure 
of the factorization theorem in Eq.~(\ref{eq:doublehemifactorization}) remains unchanged
when the quark mass is close to the soft scale $\mu_S\sim k_l\sim k_r\sim Q\tau$~\cite{Gritschacher:2013pha}. In
this case the massive quarks contribute in a nontrivial 
way only to the singular terms in the soft function $S$. So for the purpose of this
work we for the most part consider $n_f$ massless quark flavors and one additional quark flavor
with mass $m\sim\mu_S\sim Q\tau\ll Q$ in the dijet limit.
The hemisphere soft function in the factorization
theorem in Eq.~(\ref{eq:doublehemifactorization}) is defined as  
\begin{align}
 S(k_r,k_l,m,\mu)\equiv & \,  \frac{1}{N_c} \sum_{X_s}\, \langle 0 \lvert\overline{Y}_{\bar{n}}
 Y_n(0)\lvert X_s \rangle \langle X_s \lvert Y_n^{\dagger}
 \overline{Y}_{\bar{n}}^{\dagger} (0)\lvert 0\rangle \nn \\
 & \times \delta(k_l-\bar{n}\cdot k_s^l)\, \delta(k_r-n\cdot k_s^r)\, , 
\label{eq:softftc_def}
\end{align}
where $k_s^l$ ($k_s^r$) is the light-cone momentum of the soft final state
$\lvert X_s \rangle$ in the left (right) hemisphere and $Y_n(x)$,
$\overline{Y}_{\bar{n}}(x)$ are ultrasoft Wilson lines, i.e.\ 
\begin{align}
  Y_n(x) &\equiv \overline{\textrm{P}} \,  \textrm{exp}\left[-ig\int_{0}^{\infty}
   ds \, n\cdot A_{us}(ns+x)\right] \, ,  \nn \\
  \overline{Y}_{\bar{n}}(x) & \equiv
 \overline{\textrm{P}} \,  \textrm{exp}\left[-ig\int_{0}^{\infty} ds \, \bar{n}
   \cdot \overline{A}_{us}(ns+x)\right] 
\end{align}
with
$\bar{A}_{\mu}=\bar{T}^A A_{\mu}^A$ and $\bar{T}^A=-(T^A)^T$.
We have indicated the dependence of the soft function on the quark mass by the
additional argument $m$. 
For purely massless quarks the ${\cal O}(\alpha_s^2)$ corrections to the
partonic hemisphere soft function have been computed in 
Ref.~\cite{Kelley:2011ng}, see also Ref.~\cite{Hornig:2011iu}. In the following, we
will calculate the ${\cal O}(\alpha_s^2 C_F T_F)$ corrections from massive
quarks due to gluon splitting as shown in the diagrams of
Fig.~\ref{softfunction_diagrams}. The
generalization to several massive quark flavors is straightforward. 
We emphasize that the definition of the soft function in
Eq.~(\ref{eq:softftc_def}) also applies for the case when the primary quarks
are massive~\cite{Fleming:2007qr,Fleming:2007xt} as long as the c.m.\ energy $Q$ is much larger than their mass since the
emergence of the Wilson lines in the soft Eikonal approximation is
mass-independent. So our result 
also applies for massive secondary quark effects in massive primary quark
production, where masses of primary and secondary quarks can differ, up to
potential trivial modifications concerning the scheme for $\alpha_s$. 

\section{Outline of the calculation}\label{sec:outline}

\begin{figure*}
  \centering
  \includegraphics[width=13cm]{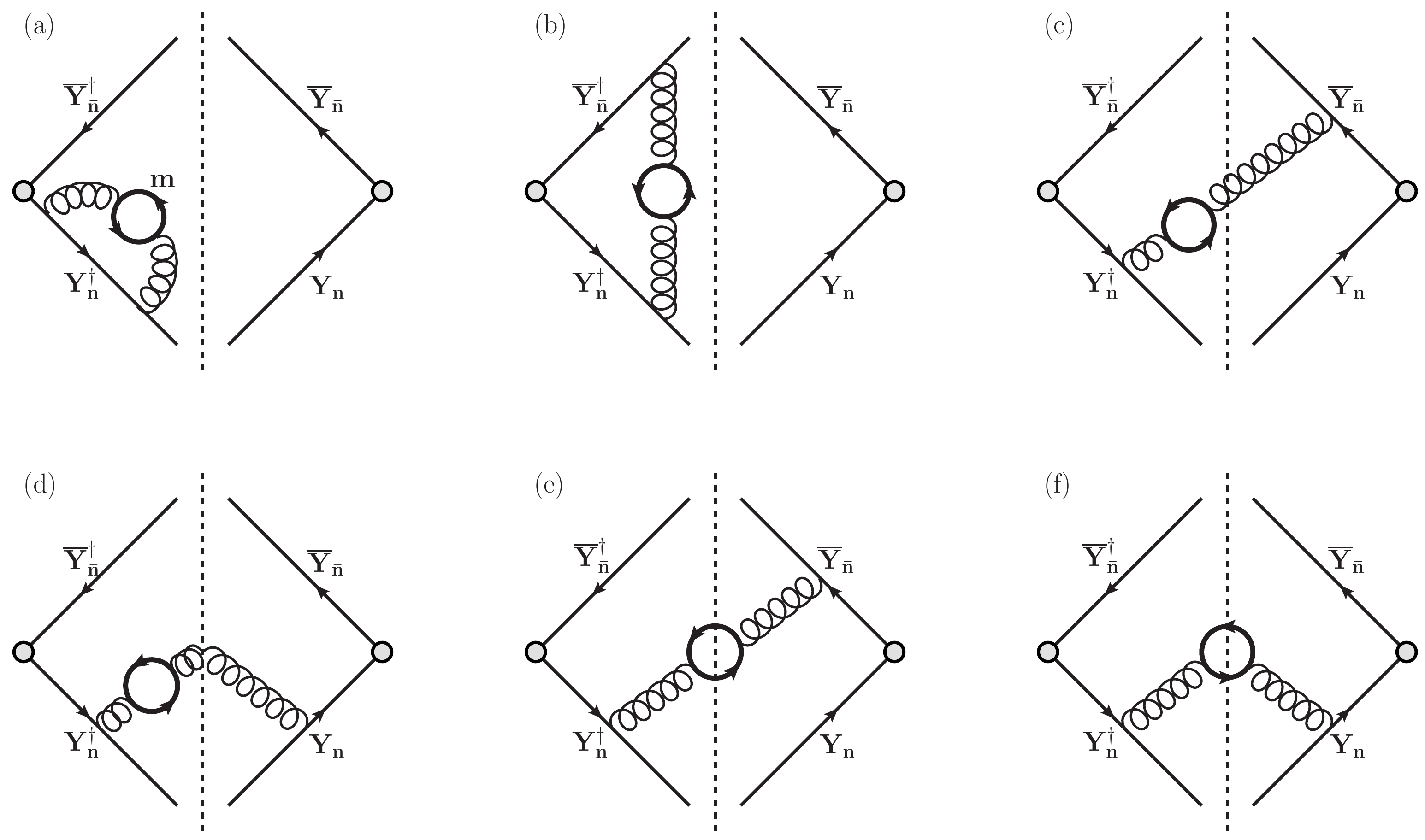}
  \caption{The types of Feynman diagrams for the $\mathcal{O}(\alpha_s^2 C_F T_F)$ contributions to the soft function with corresponding symmetric configurations. (a) and (b) are purely virtual, (c) and (d) contain the contributions to real gluon radiation and (e) and (f) the contributions for the real radiation of a quark-antiquark pair. (a) and (d) vanish due to $n\cdot n=0$.}
  \label{softfunction_diagrams}
 \end{figure*}

We compute the diagrams in Fig.~\ref{softfunction_diagrams} for one
quark flavor with mass $m$. While the phase space for the diagrams (a) - (d)
is easy, the computation of diagrams (e) and (f) is non-trivial, when the
massive quark and antiquark with momenta $k$ and $q$,
respectively, enter the final  
state. Taking into account the corresponding symmetry factors, one can obtain their
contributions to the soft function in the form analogously to the massless case~\cite{Kelley:2011ng}
\begin{align}\label{eq:S_ef}
 S_{e+f}(k_r,k_l,m)= &\int \frac{d^d k}{(2\pi)^d} \int \frac{d^d q}{(2\pi)^d}  \,
 F^{(qq)}(k,q,k_r,k_l,m) \nn \\
 & \times s(k,q) \, , 
\end{align}
where $s(k,q)$ is the matrix element calculated by conventional Feynman rules,
\begin{align}
 s(k,q)=g^4C_F T_F  \tilde{\mu}^{2\epsilon}\frac{4(k^{+}q^{-}+k^{-}q^{+}-2k\cdot
   q)}{(k^{+}+q^{+})(k^{-}+q^{-})(k+q)^4} \, . 
\end{align}
The phase space constraints and the on-shell condition for the massive quarks
are given by the \emph{quark hemisphere prescription} $F^{(qq)}(k,q,k_r,k_l,m)$,
\begin{align}
 &F^{(qq)}(k,q,k_r,k_l,m) =  \nn \\
 & (-2\pi i)^2 \, \delta(k^2-m^2) \, \delta(q^2-m^2)\,  \theta(k^{+}+k^{-}) \,  \theta(q^{+}+q^{-}) \nn \\
 &\times \left[\theta(k^{+}-k^{-}) \, \theta(q^{-}-q^{+}) \, \delta(q^{+}-k_r) \, \delta(k^{-}-k_l) \right. \nn \\
 & \hspace{0.2cm}+ \theta(k^{-}-k^{+}) \, \theta(q^{+}-q^{-}) \, \delta(k^{+}-k_r) \, \delta(q^{-}-k_l)  \nn \\
 & \hspace{0.2cm}+ \theta(k^{-}-k^{+}) \, \theta(q^{-}-q^{+}) \, \delta(k^{+}+q^{+}-k_r) \, \delta(k_l) \nn \\
 & \hspace{0.2cm}+ \left. \theta(k^{+}-k^{-}) \, \theta(q^{+}-q^{-}) \, \delta(k^{-}+q^{-}-k_l) \, \delta(k_r) \right] \, .
\end{align}
Solving the integral~(\ref{eq:S_ef}) directly with this phase space constraint turns out to 
be an extraordinarily difficult task due to the mass dependence together with the complications that arise from the
parts of the phase space where the quark and antiquark enter different
hemispheres. Instead of approaching with brute force, we 
therefore apply the following strategy: We first calculate a soft function with
a much simpler phase space constraint (but the same matrix element), 
\begin{align}\label{eq:Sg_ef}
 S_{e+f}^{(g)}(k_r,k_l,m)=&\int \frac{d^d k}{(2\pi)^d} \int \frac{d^d q}{(2\pi)^d}
 \, F^{(g)}(k,q,k_r,k_l,m) \nn \\
 & \times s(k,q) \, , 
\end{align}
where the phase space constraints are given by the \emph{gluon hemisphere
  prescription} $F^{(g)}(k,q,k_r,k_l,m)$, 
\begin{align}
 & F^{(g)}(k,q,k_r,k_l,m) = \nn \\
 & (-2\pi i)^2 \, \delta(k^2-m^2) \, \delta(q^2-m^2)\,  \theta(k^{+}+k^{-}) \,  \theta(q^{+}+q^{-}) \nn \\
 &\times \left[\theta(k^{+}+q^{+}-k^{-}-q^{-})\, \delta(k_r) \, \delta(k^{-}+q^{-}-k_l)  \right. \nn \\
 & \hspace{0.2cm}+ \left. \theta(k^{-}+q^{-}-k^{+}-q^{+})\, \delta(k_l) \, \delta(k^{+}+q^{+}-k_r)\right] \, .
\end{align}
This phase space assigns the soft hemisphere momenta coherently to the
components of the gluon momentum $k+q$, so that the massive quark and antiquark momenta
always contribute together and homogeneously to $k_l$ and $k_r$.
The soft function obtained in this way only keeps track of the
hemisphere, into which the virtual gluon propagated, and therefore differs
from the actual physical hemisphere soft function we aim to calculate, where the final
state partons each are accounted for in the hemisphere they propagate. Since both prescriptions are
compatible with soft-collinear factorization and lead to the same hard current
and jet functions, the consistency of the renormalization group evolution 
forces both soft functions to have the same UV divergences. So the required
additional correction arising from the difference between the quark hemisphere and gluon hemisphere prescription, which we call {\it phase space misalignment correction}, can be computed in four dimensions, which can be tackled numerically. Due to the finite quark masses the
resulting calculations are also IR-finite and straightforward to carry out.

The advantage of introducing the gluon hemisphere prescription is that the
kinematics is only governed by the gluon momenta weighted by the gluon
virtuality. So in the calculation the physical effects associated to the fact
that a massive quark pair is produced from virtual gluon decay can be separated
from the computation of the phase space. This makes the gluon hemisphere prescription 
quite simple to compute because it allows us to perform the computation with
the help of dispersion integrations over the gluon virtuality as described in
Refs.~\cite{Kniehl:1988id,Hoang:1994it,Hoang:1995ex}\footnote{ 
The dispersion method is actually well known from numerous previous multi-loop
calculations and renormalon studies, as well as in phenomenological applications
such as the hadronic contributions to $g-2$.
}: As a first step one calculates the ${\cal O}(\alpha_s)$ corrections to the
partonic soft function coming from the radiation of a ``massive gluon'' with momentum  
$p=k+q$. Then, by convoluting the massive gluon result with the imaginary part of the
gluon vacuum polarization function related to the massive quark cuts in diagrams
(e) and (f) one obtains the ${\cal O}(\alpha_s^2 C_F T_F)$ massive quark
corrections in the gluon hemisphere prescription. The calculation is very
generic and it is trivial to determine the effects of gluon splitting into any
other kind of final state, such as gluino pairs, just to mention one example. Note
that the method applies regardless of whether the physical effects are related to
virtual corrections or real radiation final states.

To explain the dispersion method for an equal-mass quark-antiquark pair we start with  
the gluonic vacuum polarization $\Pi(m^2,p^2)$ contribution arising from a
massive quark-antiquark bubble, 
\begin{align}
& -i\left(p^2 g_{\mu \nu}-p_{\mu} p_{\nu}\right) \Pi(m^2,p^2) 
\,\delta^{AB} \nn \\
& \equiv \int d^4 x \,e^{i p x}\, \langle 0\lvert T\left[J_{\mu}^A(x)J_{\nu}^B(0)\right] \lvert 0 \rangle \, ,
\end{align}
with the current $J^A_\mu(x)=ig_s\bar{q}(x)T^A\gamma_{\mu}q(x)$, which can be expressed through an integral over
its absorptive part. The unsubtracted (unrenormalized) dispersion integral reads  
\begin{equation}
\Pi(m^2,p^2) = -
\frac{1}{\pi} \int{dM^2 \, \frac{\mathrm{Im} \left[\Pi(m^2,M^2)\right]}{p^2-M^2+i \epsilon}
} \, ,
\label{eq:dispersion1unsubtracted}
\end{equation}
where the absorptive part in $d$ dimensions reads
\begin{align}\label{eq:Im_Pi}
&\mathrm{Im}\left[\Pi(m^2,p^2)\right] = \theta(p^2-4m^2) \, g^2 T_F \tilde{\mu}^{2\epsilon}(p^2)^{(d-4)/2} \nn \\
&\times \frac{2^{3-2d}\pi^{(3-d)/2}}{\Gamma\left(\frac{d+1}{2}\right)} \left(d-2+\frac{4m^2}{p^2}\right)\left(1-\frac{4m^2}{p^2}\right)^{(d-3)/2} \, .
\end{align}
We call this dispersion relation ``unrenormalized'' because it is related to the
calculation where the strong coupling is still unrenormalized (with respect to
the effects of the massive quark flavor). At this point the standard scheme
choices for the renormalization of the strong coupling are the $\overline {\rm MS}$
scheme involving the subtraction of the $1/\epsilon$ divergence in $\Pi(m^2,p^2)$ or
the on-shell subtraction scheme involving the subtraction of $\Pi(m^2,p^2=0)$. Using
the $\overline {\rm MS}$ scheme means that the massive quark is active concerning the
renormalization group evolution, so the strong coupling evolves with $n_f+1$
active dynamical flavors. Using on-shell subtractions means that that the
massive quark is not active concerning the renormalization group, so the strong
coupling evolves with $n_f$ active dynamical flavors. The on-shell
subtraction is used when the massive quark is integrated out and can also be implemented into the dispersion relation itself by
employing its subtracted form
\begin{align}
\Pi^{\rm OS}(m^2,p^2)& = 
\Pi(m^2,p^2)-\Pi(m^2,0) \nn \\
&=-\frac{p^2}{\pi} \int {\frac{dM^2}{M^2} \, \frac{\mathrm{Im} \left[\Pi(m^2,M^2)\right]}{p^2-M^2+i \epsilon}
  } \, .
\label{eq:dispersion1subtracted}
\end{align}
The subtracted dispersion relation has an important computational advantage
since the integration over the virtual gluon mass is suppressed by an additional
power of $1/M^2$ for large values of $M^2$. This property can make the dispersion integration
finite and may allow us to carry out the integral directly in $d=4$ dimensions. The unrenormalized result can then be easily
recovered by adding back the massless gluon result times $\Pi(m^2,p^2=0)$. We will use this method in the following.

To be definite, the correction to the Feynman gauge gluon propagator due to a massive
quark-antiquark loop expressed in terms of the subtracted
dispersion relation reads
\begin{align}
&\Pi_{\mu\nu}^{\textnormal{eff,OS}}(m^2,p^2)\equiv \,\frac{(-i)^2
  g_{\mu\rho}\Pi^{\rho\sigma,{\rm OS}}(m^2,p^2)\,g_{\sigma\nu}}{(p^2+i
  \epsilon)^2} \nn \\
  &=\,\frac{1}{\pi} \int\frac{dM^2}{M^2}\,\frac{-i\left(g_{\mu\nu}-\frac{p_\mu
        p_\nu}{p^2}\right)}{p^2-M^2+i \epsilon} \,\mathrm{Im} \left[\Pi(m^2,M^2)\right] \,, 
\label{eq:effectivegluonpropagatorsubtracted}
\end{align}
where $p^\mu$ denotes the external gluon momentum, and we have dropped from the
equalities the overall color conserving Kronecker $\delta^{AB}$. 
Note that in Eq.~(\ref{eq:effectivegluonpropagatorsubtracted}) the propagator
becomes transverse from the insertion of the vacuum polarization. 
In our calculations the contributions from the additional $p^\mu p^\nu$ term
vanish due to gauge invariance and can be ignored. The relation also shows
explicitly that we can obtain the result for the massive quark-antiquark pair
from a dispersion integral over the corresponding result for a gluon with mass $M$. 
Thus we can obtain the ${\cal O}(\alpha_s^2 C_FT_F)$-corrections to the soft
function in the gluon hemisphere prescription with the on-shell subtraction for
the strong coupling by
the relation
\begin{align}
S_{\rm{OS}}^{(g)}(k_l,k_r,m,\mu) = &\, \frac{1}{\pi} \int
\frac{dM^2}{M^2} \, S^{(1)}_M(k_l,k_r,M,\mu) \nn \\
& \times \mathrm{Im} \left[\Pi(m^2,M^2)\right] \, ,
\label{eq:dispersion1soft}
\end{align}
where $S^{(1)}_M$ denotes the one-loop massive gluon contribution to the soft function.
The corresponding result with the more common $\overline {\rm MS}$ subtraction then
reads 
\begin{align}
& S_{\rm{\overline{MS}}}^{(g)}(k_l,k_r,m,\mu) =  S_{\rm{OS}}^{(g)}(k_l,k_r,m,\mu) \nn \\
& - \left(\Pi(m^2,0)-\frac{\alpha_s T_F}{3\pi} \,\frac{1}{\epsilon}\right) \times S^{(1)}(k_l,k_r,\mu)\, 
\label{eq:dispersionMSbar}
\end{align}
with $S^{(1)}$ being the massless one-loop contribution to the soft function.
For convenience we also give the result for the zero-momentum vacuum
polarization function in $d$ dimensions: 
\begin{align}
\Pi(m^2,0) =  \frac{\alpha_s T_F}{3\pi} \left(\frac{\mu^2}{m^2}\right)^{\epsilon} \Gamma(\epsilon)e^{\gamma_E \epsilon} \, .
\label{eq:vacpolzero}
\end{align}

\section{Massive quark corrections with gluon hemisphere prescription}\label{sec:gluonhemisphere}

We start with the computation of ${\cal O}(\alpha_s^2 C_F T_F)$ massive quark
corrections to the soft function with the gluon hemisphere prescription,
$S^{(g)}(k_r,k_l,m,\mu)$, along the lines described above. 

\subsubsection{Soft function for massive gluons at ${\cal O}(\alpha_s)$}

In the calculations for the ${\cal O}(\alpha_s)$ soft function with a massive
gluon, we encounter rapidity divergences in individual parts of the computation
which are not regularized by dimensional regularization. Although the rapidity
divergences cancel in the sum of all terms~\cite{Gritschacher:2013pha}, it is convenient to
implement an additional regulator. In any case, using a regulator, all integrals
are well defined individually, and the outcome for the different diagrams is
regulator-dependent. 
We choose the ``$\alpha$-regulator''
\cite{Smirnov:1997gx,Becher:2011dz}
\begin{equation}
 \int dp^{-} \rightarrow \int dp^{-} \left(\frac{\nu}{p^{-}}\right)^{\alpha} \,
 \label{alpharegulator}
\end{equation}
on the minus gluon momentum component. We use the $\alpha$-regulator in a way more general than advocated in
Ref.~\cite{Becher:2011dz} since we apply it not only for phase space integrations but also for loop diagrams, so that some of the properties of this regulator for phase space
integrals as stated in Ref.~\cite{Becher:2011dz} might not hold. The exact implementation of the regularization of rapidity
divergences is only of minor importance for our calculation since the 
divergences cancel entirely within the soft function computation and no large logarithms arise when the mass is of order of the soft scale.
With the regulator~(\ref{alpharegulator}) the purely virtual diagram
in Fig.~\ref{softfunction_diagrams} (b) yields a 
scaleless integral and vanishes, so only the diagrams containing the real radiation of a
massive gluon are non-vanishing.\footnote{We emphasize that this is a regulator dependent statement.}

 The diagrams for real radiation of a gluon with mass $M$ in 
$d$-dimensions, after expanding in $\alpha$, yield ($\bar{k}=k/\mu$)
\begin{align}\label{soft_d}
  & \mu^2 S^{(1)}_M (k_r,k_l,M,\mu) = \frac{\alpha_s(\mu) C_F}{4\pi}\,\delta(\bar{k}_l)
  \left\{\left(\frac{\mu^2}{M^2}\right)^\epsilon e^{\gamma_E \epsilon}
    \Gamma(\epsilon) \right. \nn \\
   & \times \left(2\,\delta(\bar{k}_r)\left[\ln\left(\frac{\mu^2}{M^2}\right)+\gamma_E-\psi(\epsilon)\right] +4\left[\frac{\theta(\bar{k}_r)}{\bar{k}_r}\right]_{+} \right)\nn\\
   &\left. -\,\theta(k_r-M)\frac{4}{\bar{k}_r} \, \ln\left(\frac{k_r^2}{M^2}\right) \right\}  + (k_r \leftrightarrow k_l) \, . 
\end{align}
Note that the threshold term involving the $\theta$-function corresponds to real
radiation. It has been given for $d=4$ ($\epsilon=0$) because it only involves
IR and UV finite integrals within the subtracted dispersion
integral~(\ref{eq:dispersion1soft}).  
Technical details on the calculation leading to Eq.~(\ref{soft_d}) can be found
in \cite{Gritschacher:2013pha}. 

\subsubsection{Massive quark corrections at ${\cal O}(\alpha_s^2)$}

We use the dispersive technique to obtain the ${\cal O}(\alpha_s^2 C_F T_F)$ part of the soft
function for massive quarks. The convolution along the lines of
Eq.~(\ref{eq:dispersion1soft}) is performed separately for the $d$
dimensional virtual terms and the four-dimensional threshold term in Eq.~(\ref{soft_d}),
where for the latter the $d=4$ version of the absorptive part of the vacuum
polarization function in Eq.~(\ref{eq:Im_Pi}) can be used. We encounter
hypergeometric functions, which are expanded with the $HypExp$ package~\cite{Huber:2005yg}
in $Mathematica$. The structure of the result with on-shell subtraction
for the strong coupling ($\alpha_s=\alpha_s^{(n_f)}$) reads  
\begin{align}
 S^{(g)}_{OS}(k_l,k_r,m,\mu)= S^{(g)}_{OS,\rm virt}(k_l,k_r,m,\mu)+S^{(g)}_{\rm
   real}(k_l,k_r,m) \, , 
\end{align}
where the unrenormalized distributive part $S^{(g)}_{OS,\rm virt}$ describing
virtual radiation reads 
\begin{align}
 & \mu^2 S^{(g)}_{OS,\rm virt}(k_l,k_r,m,\mu)= \frac{\alpha_s^2 C_F T_F}{16\pi^2} \,\delta(\bar{k}_l) \bigg\{\delta(\bar{k}_r) \nn \\
 & \times \left[-\frac{2}{\epsilon^3}+\frac{1}{\epsilon^2}\left(\frac{8}{3}L_m+\frac{10}{9}\right)+\frac{1}{\epsilon}\left(-\frac{4}{3}L_m^2 +\frac{56}{27}-\frac{\pi^2}{3}\right) \right. \nn \\
 & -\left. \frac{20}{9}L_m^2+\left(-\frac{224}{27}+\frac{4\pi^2}{9}\right)L_m-\frac{328}{27}+\frac{5\pi^2}{27}+4\zeta(3) \right] \nn \\
 & + \left[\frac{\theta(\bar{k}_r)}{\bar{k}_r}\right]_{+}\left[\frac{8}{3\epsilon^2}+\frac{1}{\epsilon}\left(-\frac{16}{3}L_m-\frac{40}{9}\right)+\frac{16}{3}L_m^2+\frac{80}{9}L_m \right.\nn\\
 & + \left.\left.\frac{224}{27}+\frac{4\pi^2}{9}\right]\right\} +  (k_r \leftrightarrow k_l) \, ,
\end{align}
with $L_m=\ln(m^2/\mu^2)$. The finite threshold part $S^{(g)}_{\rm
  real}(k_l,k_r,m,\mu)$ describing real radiation and vanishing for $k_{l,r}
\leq 2m$ reads 
\begin{align}\label{eq:soft_gluon_massive_real}
 & \mu^2 S^{(g)}_{\rm real}(k_l,k_r,m) =\frac{\alpha_s^2 C_F T_F}{16\pi^2} \,\delta(\bar{k}_l) \,\theta(k_r-2m) \nn \\
 & \times\frac{1}{\bar{k}_r} R\left(\sqrt{1-\frac{4m^2}{k_r^2}}\right) +  (k_r \leftrightarrow k_l)\, ,
\end{align}
where the real radiation function is given by
\begin{align}\label{eq:Rfunction}
 & R(w)=  \frac{32}{3} \Li_2\left(\frac{w-1}{w+1}\right)+\frac{16}{3}\ln^2 \left(\frac{1+w}{2}\right) \nn \\
 &- \frac{16}{3}\ln^2 \left(\frac{1-w}{2}\right) +\frac{8}{3}\ln^2 \left(\frac{1+w}{1-w}\right)+\frac{80}{9} \ln \left(\frac{1+w}{1-w}\right) \nn \\
 & + \frac{32}{27}w^3-\frac{160}{9}w+\frac{8\pi^2}{9} \, .
\end{align} 
To obtain the result with $\overline {\rm MS}$ subtraction for the strong
coupling ($\alpha_s=\alpha_s^{(n_f+1)}$) one has to add the $\MS$-renormalized
finite contributions
of the zero momentum vacuum polarization according to
Eq.~(\ref{eq:dispersionMSbar}). This gives the additional contributions 
\begin{align}\label{eq:deltaS_dispersion}
 &\delta S^{(g)}(k_r,k_l,m,\mu) =-\left(\Pi(m^2,0)-\frac{1}{\epsilon}\right) \times S^{(1)}(k_r,k_l,\mu) \nn \\ 
 &=-\frac{\alpha_s T_F}{3\pi}\left[\left(\frac{\mu^2}{m^2}\right)^{\epsilon}\Gamma(\epsilon)e^{\gamma_E
     \epsilon}-\frac{1}{\epsilon}\right]\times \frac{\alpha_s C_F}{\pi}\frac{\mu^{2\epsilon} e^{\gamma_E \epsilon}}{\epsilon \Gamma(1-\epsilon)} \nn \\
     & \hspace{0.4cm}\times \left[\delta(k_l)\theta(k_r)k_r^{-1-2\epsilon} + (k_r \leftrightarrow k_l)\right] \, . 
\end{align}
The contributions in Eq.~(\ref{eq:deltaS_dispersion}) contain only the virtual
distributive pieces and do not affect the massive quark real radiation
corrections. Expanded for small $\epsilon$ the unrenormalized ${\cal
  O}(\alpha_s^2 C_F 
T_F)$ massive quark contributions  to the soft function (with  $\overline {\rm MS}$ renormalized
$\alpha_s^{(n_f+1)}$) read
\begin{align}\label{eq:soft_gluon_massive}
 S^{(g)}(k_r,k_l,m,\mu)= & \, S^{(g)}_{OS}(k_l,k_r,m,\mu)+\delta S^{(g)}(k_r,k_l,m,\mu) \nn \\
 = & \, Z_{S,m}(k_r,k_l,\mu) + S^{(g)}_{\rm virt}(k_r,k_l,m,\mu) \nn \\
 & + S^{(g)}_{\rm real}(k_r,k_l,m,\mu) \, ,
\end{align} 
where the UV-finite contribution to the distributive virtual corrections is given by ($L_m=\ln{(m^2/\mu^2)}$)
\begin{align}\label{eq:soft_gluon_massive_virt}
& \mu^2 S^{(g)}_{\rm virt}(k_r,k_l,m,\mu)= \frac{\alpha_s^2 C_F T_F}{16\pi^2} \,\delta(\bar{k}_l)\left\{\delta(\bar{k}_r)\left[-\frac{4}{9}L_m^3\right.\right.\nn\\
& -\left.\frac{20}{9}L_m^2+\left(-\frac{224}{27}+\frac{4\pi^2}{9}\right)L_m-\frac{328}{27}+\frac{5\pi^2}{27}+\frac{28}{9}\zeta(3) \right] \nn \\
 & + \left[\frac{\theta(\bar{k}_r)}{\bar{k}_r}\right]_{+}\left[\frac{8}{3}L_m^2+\frac{80}{9}L_m+\frac{224}{27}\right] \nn \\
 & - \left.\left[\frac{\theta(\bar{k}_r)\ln \, \bar{k}_r}{\bar{k}_r}\right]_{+} \frac{32}{3}L_m  \right\} + (k_r \leftrightarrow k_l) \, ,
\end{align} 
and the UV-divergent contribution, which gives the $\mathcal{O}(\alpha_s C_F T_F)$ massive quark correction to the soft function renormalization factor, reads
\begin{align}\label{eq:Z_Sm}
 & \mu^2 Z_{S,m}(k_r,k_l,\mu)= \frac{\alpha_s^2 C_F T_F}{16\pi^2} \,\delta(\bar{k}_l)\left\{\delta(\bar{k}_r)\left[-\frac{2}{\epsilon^3}+\frac{10}{9\epsilon^2}\right.\right. \nn\\
 & \left.+\left.\frac{1}{\epsilon}\left(\frac{56}{27}-\frac{\pi^2}{9}\right)\right] + \left[\frac{\theta(\bar{k}_r)}{\bar{k}_r}\right]_{+}\left[\frac{8}{3\epsilon^2}-\frac{40}{9\epsilon}\right] \right\} + (k_r \leftrightarrow k_l) \, .
\end{align}
The UV divergences agree exactly with the known result for one massless quark, since the mass is just an infrared scale and does not affect the UV behavior. 
Therefore, the secondary massive quark flavor contributes to the anomalous dimension of the soft function in the same way as a massless flavor.


From
Eq.~(\ref{eq:soft_gluon_massive}) one can take the massless limit by expanding
the real radiation contribution $S^{(g)}_{\rm real}$ into delta functions and
plus distribution, which leads to the (unrenormalized) result
\begin{align}\label{eq:soft_gluon_massless}
& \mu^2 {S}^{(g)} (k_r,k_l,\mu) = \mu^2 Z_S(k_r,k_l,\mu) + \frac{\alpha_s^2 C_F T_F}{16\pi^2}\,
\delta(\bar{k}_l) \nn\\
& \times \left\{\delta(\bar{k}_r) \left[\frac{328}{81}-\frac{5\pi^2}{9}-\frac{20}{9}\zeta(3) \right]+\left[\frac{\theta(\bar{k}_r)}{\bar{k}_r}\right]_{+}\left[-\frac{224}{27}\right. \right. \nn \\ 
 & \left.+ \left.\frac{8\pi^2}{9}\right] +\left[\frac{\theta(\bar{k}_r)\,\ln \,
       \bar{k}_r}{\bar{k}_r}\right]_{+}\frac{160}{9} -
   \left[\frac{\theta(\bar{k}_r)\,\ln^2
       \bar{k}_r}{\bar{k}_r}\right]_{+}\frac{32}{3} \right\}  \nn \\
       & + (k_r \leftrightarrow k_l) \,.
\end{align}

\section{Phase space misalignment correction}\label{sec:quarkhemisphere}
\label{sec:Shemicomputation}

We now determine the ${\cal O}(\alpha_s^2 C_F T_F)$ massive quark corrections to
the double hemisphere soft function for the physical quark hemisphere
prescription. After having obtained the result for the gluon hemisphere
prescription ${S}^{(g)} (k_r,k_l,m,\mu)$ in Sec.~\ref{sec:gluonhemisphere}, what remains to be calculated are the
corrections due to the phase space misalignment to the physical quark hemisphere
prescription, which we call $\Delta S(k_l,k_r,m)$. So the result for the full ${\cal
  O}(\alpha_s^2 C_F T_F)$ massive quark corrections to 
the unrenormalized double hemisphere soft function reads
\begin{align}\label{eq:DeltaS_tot}
S_m(k_l, k_r,m,\mu) = S^{(g)}(k_l, k_r,m,\mu) + \Delta S(k_l,k_r,m)\,.
\end{align} 
The phase space misalignment correction $\Delta S$ contains only phase space
contributions, where the two quarks enter  
different hemispheres, since quark and gluon hemisphere prescriptions act in the
same way when the quarks enter the same hemisphere.
After having performed the integrations over the transverse momenta in
Eqs.~(\ref{eq:S_ef}) and (\ref{eq:Sg_ef}) we obtain
\begin{align}\label{eq:DeltaS_hemiPS}
 &\Delta S (k_l,k_r,m)= \frac{\alpha_s^2 C_F T_F}{16\pi^2}\int dq^{-} \int dk^{+} \int dq^{+} \int dk^{-}  \nn \\
 & \times \theta(k^{-}-k^{+})\,\theta(q^{+}-q^{-})\,\theta(k^{+}k^{-}-m^2)\,\theta(q^{+}q^{-}-m^2) \nn \\
 & \times \theta(k^{-}+k^{+})\,\theta(q^{+}+q^{-})\left[\delta(k_l-q^{-})\,\delta(k_r-k^{+})\right. \nn \\
 & -\theta(k^{-}+q^{-}-k^{+}-q^{+})\,\delta(k_r-k^{+}-q^{+})\,\delta(k_l)\nn \\
 & -\left. \theta(k^{+}+q^{+}-k^{-}-q^{-})\,\delta(k_l-k^{-}-q^{-})\,\delta(k_r)\right] \nn \\
 & \times f_m(k^{+},k^{-},q^{+},q^{-},m)
\end{align}
 with the integrand 
 \begin{align}\label{eq:fm}
&f_m(k^{+},k^{-},q^{+},q^{-},m)=
\left[\left(\frac{k^{+}q^{+}}{(q^{+}+k^{+})^2}+\frac{k^{-}q^{-}}{(q^{-}+k^{-})^2}\right)\right. \nn \\
&\times \left.\left(q^{+}k^{-}+k^{+}q^{-}\right)-\frac{4(k^{+}k^{-}-m^2)(q^{+}q^{-}-m^2)}{(q^{+}+k^{+})(q^{-}+k^{-})}\right]
\nn\\
 &\times
\frac{16}{\left[(q^{+}k^{-}+k^{+}q^{-})^2-4(k^{+}k^{-}-m^2)(q^{+}q^{-}-m^2)\right]^{3/2}}
\, .
\end{align}
In fact the results of both, quark and gluon hemisphere prescriptions entering $\Delta S$ are individually free of
UV divergences. Conceptually this is related to the consistency of
soft-collinear factorization and the exponentiation properties of the soft
function~\cite{Hoang:2007vb,Hoang:2008fs}, so that at ${\cal O}(\alpha_s^2)$ 
UV-divergent contributions depending simultaneously on both the hemisphere
variables $k_l$ and $k_r$ in a non-trivial way can only have $C_F^2$ color-structures.
For the massive quark corrections we calculate, there are also no IR
divergences for both hemisphere prescriptions individually since the mass acts as an IR regulator.\footnote{In the massless computation of
  \cite{Kelley:2011ng,Hornig:2011iu} infrared $1/\epsilon$-divergences arise for the phase space, where the two quarks enter opposite hemispheres, as well as for the one, where they enter the same hemisphere.
  These cancel in the sum.} Therefore we do not have to
employ any additional 
regularization and a numerical
computation can be easily performed. Furthermore, since these
contributions to the soft function correspond to real emission diagrams, no
non-trivial distributions are 
generated. 

\begin{figure}
  \begin{center}
  \subfigure{\epsfig{file=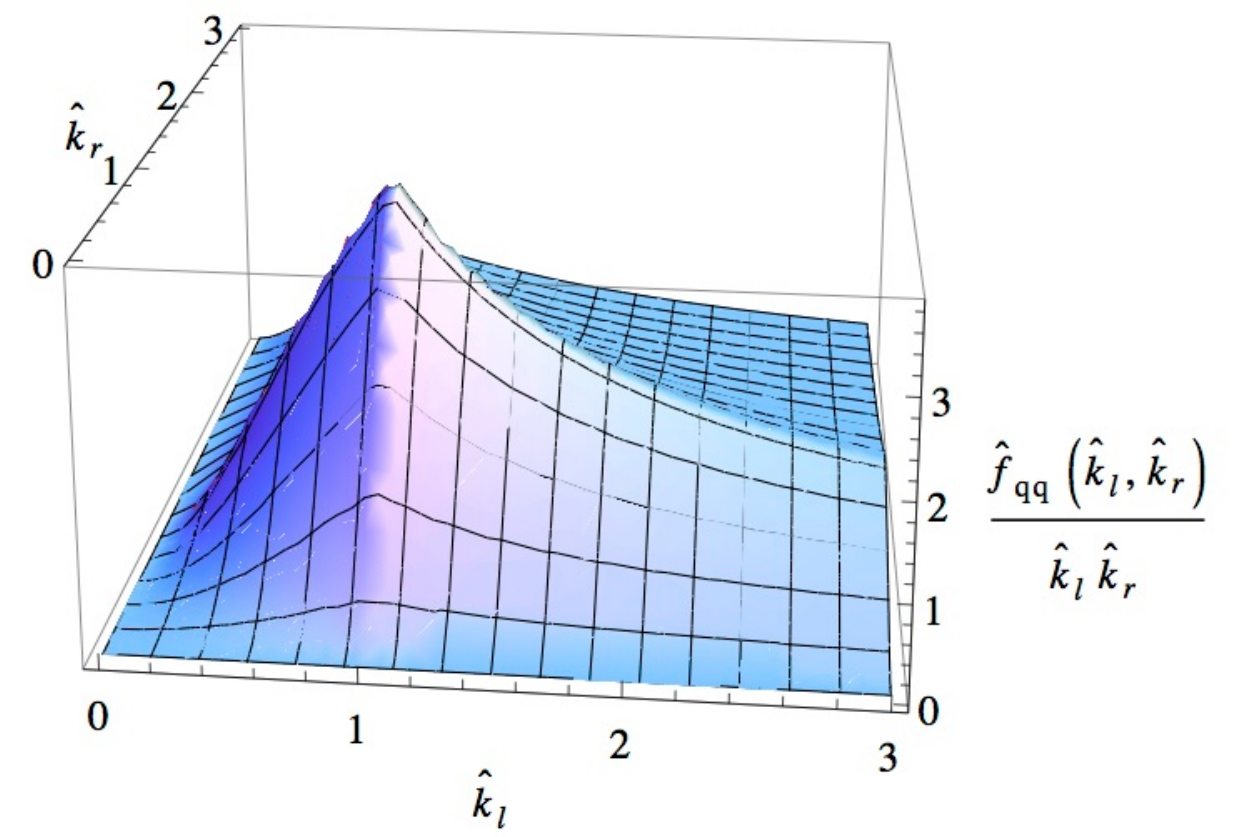,width=\linewidth,clip=}}
  \caption{3D plot of the quark hemisphere contribution to the phase space misalignment correction $\Delta S$.}
  \label{fig:f_qq_3d}
  \end{center}
\end{figure}

\begin{figure}
  \begin{center}
  \subfigure{\epsfig{file=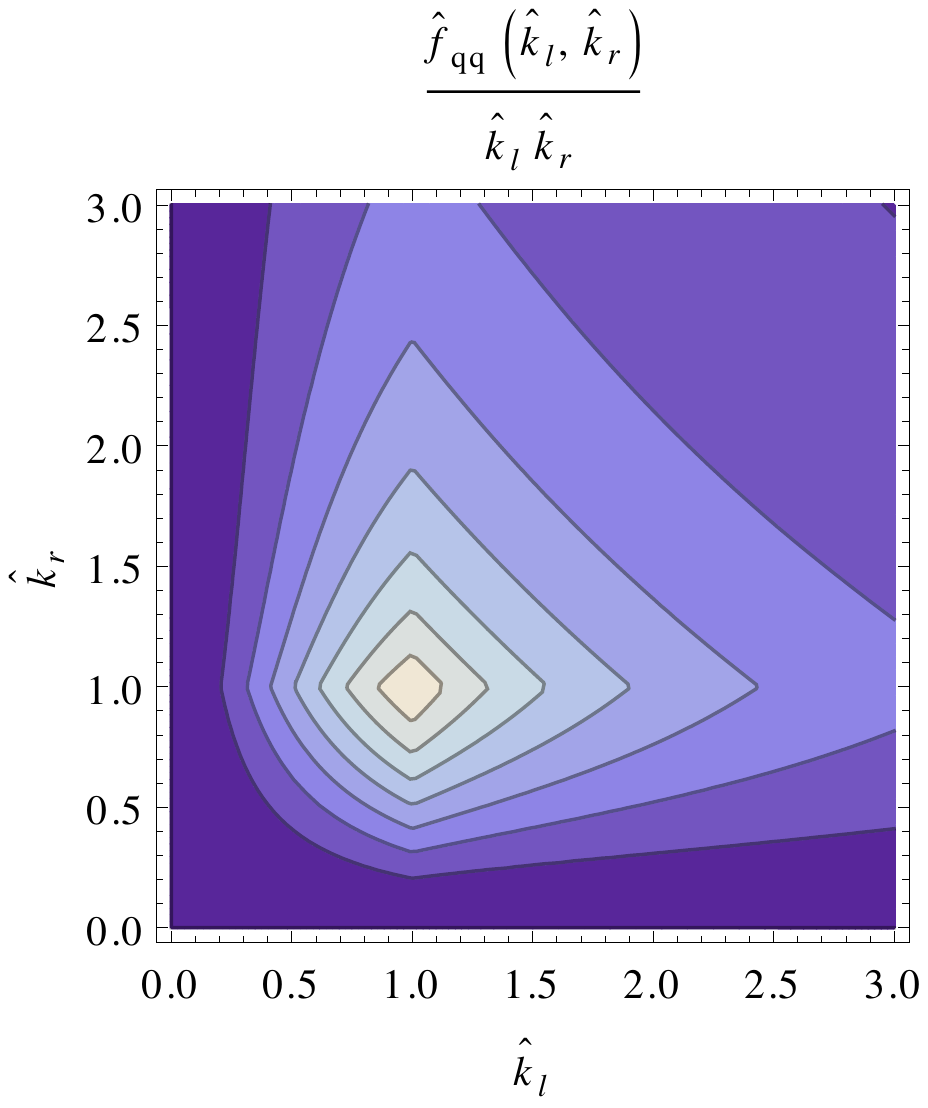,width=0.8\linewidth,clip=}}
  \caption{Contour plot of the quark hemisphere contribution to the phase space misalignment correction $\Delta S$.}
  \label{fig:f_qq_ct}
  \end{center}
\end{figure}

\begin{figure}
  \begin{center}
  \subfigure{\epsfig{file=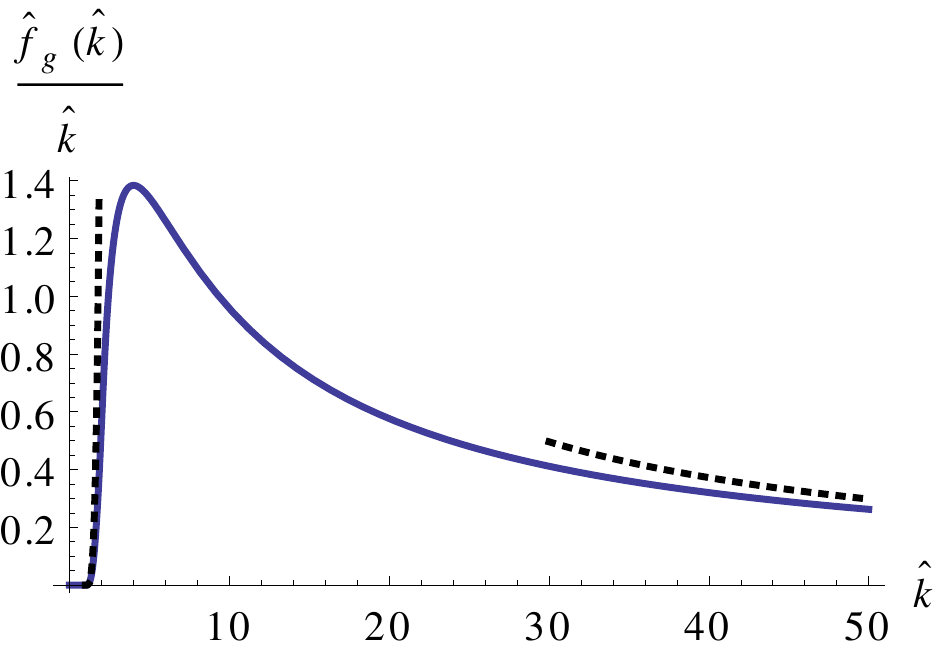,width=0.8\linewidth,clip=}}
  \caption{The gluon hemisphere contribution to the phase space misalignment correction $\Delta S$ together with its asymptotic expansions.}
  \label{fig:f_g}
  \end{center}
\end{figure}

Using Eq.~(\ref{eq:DeltaS_hemiPS}) we can cast the result for the phase space
misalignment correction into the form
($\bar{k}_{l,r}=k_{l,r}/\mu$, 
$\hat{k}_{l,r}=k_{l,r}/m$) 
\begin{align}\label{eq:Dsoftdoublenum}
 & \mu^2 \Delta {S}(k_l,k_r,m) = \frac{\alpha_s^2 C_F T_F}{16\pi^2}\left[\frac{2}{\bar{k}_l \bar{k}_r}\,\hat{f}_{qq}(\hat{k}_l,\hat{k}_r)\right. \nn \\
 &-\left. \delta(\bar{k}_l)\frac{1}{\bar{k}_r}\hat{f}_{g}(\hat{k}_r) - \delta(\bar{k}_r)\frac{1}{\bar{k}_l}\hat{f}_{g}(\hat{k}_l)\right] \, .
\end{align}
The term $\hat{f}_{qq}$ is the contribution due to the quark hemisphere prescription, 
\begin{align}\label{eq:fqq}
&\hat{f}_{qq}(\hat{k}_l,\hat{k}_r)= \frac{\hat{k}_l \hat{k}_r}{2}  \int_{0}^{\infty}
dy^{+} \int_{0}^{\infty} dx^{-} \,\theta(x^{-}-\hat{k}_r)\,\theta(y^{+}-\hat{k}_l) \nn \\
&\times \theta(\hat{k}_r x^{-}-1)\theta(\hat{k}_l y^{+}-1) f_m(\hat{k}_r,x^{-},y^{+},\hat{k}_l,1)
\end{align}
in rescaled variables with $x^{\pm}=k^{\pm}/m$ and $y^{\pm}=q^{\pm}/m$. Since $\hat{f}_{qq}$ is dimensionless we have recombined the scales and written $f_m$ as a dimensionless function in terms of these rescaled momenta. The term $\hat{f}_{g}$ is related to the gluon hemisphere prescription and reads
 \begin{align}\label{eq:fg}
& \hat{f}_{g}(\hat{k})=\hat{k} \int_{0}^{\infty} dy^-
\int_{0}^{\infty} dy^{+} \int_{0}^{\infty} dx^{-} \, \theta(x^{-}+y^{-}-\hat{k})\nn \\
&\times\theta(y^{+}-y^{-})\, \theta(\hat{k} x^{-}-y^{+}x^{-}-1)\,\theta(y^{+}y^{-}-1)\nn \\
&\times f_m(\hat{k}-y^{+},x^{-},y^{+},y^{-},1) \, .
\end{align}
Writing out the phase space in Eq.~(\ref{eq:fqq}) and (\ref{eq:fg}) in terms
of separate integration domains, one can evaluate $\hat{f}_{qq}$ and
$\hat{f}_{g}$ numerically. Using the Cuba library \cite{Hahn:2004fe} we obtained
the same result for both deterministic as well as Monte-Carlo 
algorithms. The resulting functions are displayed in the
Figs.~\ref{fig:f_qq_3d},~\ref{fig:f_qq_ct} and \ref{fig:f_g}. Notice that $\hat{f}_{qq}(\hat{k}_{l},\hat{k}_{r})$ contains a kink at $\hat{k}_{l}=1$ and $\hat{k}_{r}=1$, which can be seen in the contour plot in Fig.~\ref{fig:f_qq_ct} and can be traced back to a change of the integration domains for these values. $\hat{f}_{g}(\hat{k}_{l,r})$ contains a threshold at the scale $\hat{k}_{l,r}=1$, at which it turns on smoothly. Indeed the momentum deposit $k_{l,r}$
in the gluon prescription for the opposite hemisphere phase space is a sum of
one large lightcone component, say $q^{+}$, and a small lightcone component
$k^{+}$. Since the invariant mass of each real particle is fixed by the
on-shell condition, it follows that $q^{+}q^{-}\geq m^2$. Taking into account
that in this case $q^{+}>q^{-}$ we get $k^2_r=(q^{+}+k^{+})^2\geq (q^{+})^2 \geq
q^{+}q^{-}\geq m^2$. For $\hat{f}_{qq}$ no threshold arises, since just the small lightcone components contribute.

We can investigate the asymptotic behavior of the functions $\hat{f}_{qq}$ and
$\hat{f}_{g}$ analytically and show the results for very heavy and light quarks
explicitly. For $k_l, k_r \ll m$ the expansion of $\hat{f}_{qq}$ yields 
\begin{align}\label{eq:f_qq_MM}
\hat{f}_{qq}(\hat{k}_l,\hat{k}_r) \stackrel{\hat{k}_l,\hat{k}_r \ll 1}{\longrightarrow}  8
\hat{k}_l^3 \hat{k}_r^3 \left[1+\mathcal{O}(\hat{k}_l \hat{k}_r) \right]\, . 
\end{align}
This is the only contribution to $\Delta S$ in this limit, since $\hat f_{g}$
vanishes here, and gives the suppression of mass effects in the decoupling
limit. 
For $k_l,k_r \gg m$ we get 
\begin{align}\label{eq:f_qq_M}
\hat{f}_{qq}\left(\hat{k}_l,\hat{k}_r\right) \stackrel{\hat{k}_l,\hat{k}_r \gg 1}{\longrightarrow} f_{qq}\left(\frac{k_l}{k_r}\right)+\mathcal{O}\left(\frac{1}{\hat{k}_l},\frac{1}{\hat{k}_r}\right)
\end{align}
with
\begin{align}\label{eq:f_qq_ml}
f_{qq}(z) = \frac{16}{3}\,\ln(1+z)-\frac{8z(3+3z+2z^2)}{3(1+z)^3}\,\ln\,z-\frac{16z}{3(1+z)^2} \, .
\end{align}
Note that $f_{qq}(1/z)=f_{qq}(z)$ and $f_{qq}(0)=f_{qq}(\infty)=0$.
Turning to $\hat{f}_{g}$ we obtain for $k \gg m$ 
\begin{align}\label{eq:f_g_ml}
 \hat{f}_{g}(\hat{k}) \stackrel{\hat{k} \gg 1}{\longrightarrow}C_g \equiv -\frac{8}{3}+\frac{16\pi^2}{9}+\mathcal{O}\left(\frac{1}{\hat{k}}\right) \, .
\end{align}
So the massless limit for the misalignment correction $\Delta S(k_l,k_r,m)$ for
non-vanishing $k_l, k_r$ reads
\begin{align}\label{eq:Dsoftmzerov1}
& \mu^2 \Delta {S}(k_l>0,k_r>0,m=0) = \frac{\alpha_s^2 C_F T_F}{16\pi^2}\left[\frac{2}{\bar{k}_l \bar{k}_r}\,
f_{qq}\left(\frac{k_l}{k_r}\right) \right.\nn \\
& - \left. C_g \left( \delta(\bar{k}_l)\frac{1}{\bar{k}_r}
+ \delta(\bar{k}_r)\frac{1}{\bar{k}_l}\right)\,
\right] \, ,
\end{align}
The sum of Eq.~(\ref{eq:Dsoftmzerov1}) and the
gluon hemisphere contribution for a massless quark $S^{(g)}(k_l,k_r,\mu)$ given in
Eq.~(\ref{eq:soft_gluon_massless}) correctly reproduces the ${\cal
  O}(\alpha_s^2 C_F T_F n_f)$ massless quark corrections to the hemisphere soft
function computed in Ref.~\cite{Kelley:2011ng} for $k_l,k_r>0$. 
Note that the naive massless limits we obtain for
$\hat{f}_{qq}(\hat{k}_l,\hat{k}_r)/\hat{k}_l \hat{k}_r$ and 
$\hat{f}_{g}(\hat{k})/\hat{k}$ when  $k_l, k_r$ are non-vanishing are not integrable
at $k_l=k_r=0$. In the following section we will show, 
how they recombine into unambiguous distributive expressions.

\section{The massless limit of the hemisphere soft function}\label{sec:massless}

We now investigate the distributive structure of the ${\cal
  O}(\alpha_s^2)$ momentum-space double hemisphere 
soft function $S(k_l,k_r,\mu)$
in the massless limit. This issue has not been fully resolved
in Refs.~\cite{Kelley:2011ng,Hornig:2011iu} for the
phase space contributions where the quark and antiquark enter different
hemispheres. We will find a definite answer for 
analytic test functions $g(k_l,k_r)\neq g(k_l/k_r)$, which have in particular a
converging Taylor series around the origin.\footnote{
A test function $g(k_l,k_r)= g(k_l/k_r)$ is in general not unique at $k_l=k_r=0$ and cannot be
expanded in a Taylor series around the origin. The derivation of the distributive structure is more complicated in this case due to possible nontrivial contributions in region (a) mentioned below, and the final results presented
here do not apply.} This is usually the case for test
functions which depend on an additional scale, as it is realized for the soft
model function $S^{\rm mod}=S^{\rm mod}(k_l/\Lambda_{QCD},k_r/\Lambda_{QCD})$
which depends intrinsically on the hadronization scale
(see also Sec.~\ref{sec:renormalon_subtraction}). The computational rules we can
identify concerning the massless limit of the ${\cal O}(\alpha_s^2 C_F T_F)$ massive
quark corrections to the soft function can be related to the corresponding
massless quark results regularized in dimensional regularization given in~\cite{Kelley:2011ng,Hornig:2011iu}. They also allow us to 
determine the complete distributive structure of the pure ${\cal O}(\alpha_s^2
C_AC_F)$ gluonic corrections of the momentum-space double hemisphere soft
function. 

\subsection{Distributive structure of the ${\cal O}(\alpha_s^2 C_F T_F n_f)$
  corrections}
\label{sec:distributionmassless}

We first analyze the distributive structure of the ${\cal O}(\alpha_s^2 C_F T_F n_f)$ 
massless quark contributions from the limit 
$m\to 0$ of the massive quark corrections $S_m$ as given in Eq.~(\ref{eq:DeltaS_tot}).
The massless limit for the gluon hemisphere term $S^{(g)}$ has already been given in
Eq.~(\ref{eq:soft_gluon_massless}), so we just have to examine the double cumulant for
the phase space misalignment correction $\Delta S$ to derive the distributive structure. Given that the test functions we
consider are unique at $k_l=k_r=0$ we can identify the distributive
structures in an unambiguous way. The double cumulant is given by 
\begin{equation}\label{eq:DeltaScumulant}
 \Delta \mathcal{S}(K_L,K_R,m)= \int_0^{K_L} dk_l \int_0^{K_R} dk_r \, \Delta S(k_l,k_r,m)
\end{equation}
To reach the massless limit we perform an asymptotic expansion for $K_{L,R} \gg
m$ both for the quark and gluon hemisphere prescription contributions contained
in $\Delta S$. There are 
in principle many different relevant kinematic regimes for the lightcone components
that can contribute. Investigating the integrand $f_m(k,q,m)$ given in Eq.~(\ref{eq:fm})
and the integration measures for the lightcone components, we find only two
relevant regions giving leading $\mathcal{O}(1)$ contributions: 
(a) $k^{+} \sim k^{-} \sim q^{+} \sim q^{-} \sim m$ and 
(b) $k^{+} \sim k^{-} \sim q^{+} \sim q^{-} \sim K_{L,R}$.\footnote{Other
  conceivable regions always give a suppression of at least 
  $\mathcal{O}(m/K_{L,R})$ due to the integration measure or the power
  counting of the integrand.} 
The integration in region (a) appears to be very difficult for an analytic computation, since no expansion
is possible for the integrand $f_m(k,q,m)$. However, we can take advantage of
the fact that the phase space constraints for the gluon and quark hemisphere
prescriptions become identical in region (a). 
This can be seen from Eq.~(\ref{eq:DeltaS_hemiPS}), where after integrating in
$k_l$ and $k_r$ the dependence on the large scales $K_L$ and $K_R$ drops out for
small lightcone momenta, which leads to the cancellation between both hemisphere prescriptions.  It is
therefore sufficient to investigate only the contributions from region (b), where the
mass dependence drops from both the integrand and the domain of integration, so
that  
\begin{align}
 & \Delta \mathcal{S}(K_L,K_R,m\to 0)= \frac{\alpha_s^2 C_F T_F}{16\pi^2} \int_0^\infty dq^{-}\int_0^\infty dk^{+} \nn \\
 & \times \int_0^\infty dq^{+}\int_0^\infty dk^{-} \, \theta(q^{+}-q^{-})\,\theta(k^{-}-k^{+})\nn \\
 & \times \left[\theta(K_R-q^{-})\,\theta(K_L-k^{+})\right.\nn \\  
 &-\theta(k^{+}+q^{+}-k^{-}-q^{-})\,\theta(K_R-q^{-}-k^{-})\,\theta(K_L) \nn \\  
 &-\left. \theta(k^{-}+q^{-}-k^{+}-q^{+})\,\theta(K_L-q^{+}-k^{+})\,\theta(K_R)\right]\nn \\
 &\times f_0(k^{+},k^{-},q^{+},q^{-})\, , 
\end{align}
with the massless integrand
\begin{align}
 f_{0}(k^{+},k^{-},q^{+},q^{-})=& \frac{16}{(k^{+}+q^{+})^2 (k^{-}+q^{-})^2} \nn \\
 & \times \frac{q^{+}q^{-}+k^{+}k^{-}}{q^{+}k^{-}-k^{+}q^{-}} \, .
\end{align}
The remaining integrations can be performed separately for the contributions from the quark and gluon hemisphere prescriptions with an additional IR
regulator, where the IR divergence comes from the region where
all momenta are small. We have chosen a cutoff regulator for one of the
lightcone components and observed that it properly cancels in the final
expression in the difference between the two hemisphere contributions. This cancellation takes place since the IR divergences
in the quark and gluon hemisphere prescriptions match.\footnote{IR divergences in region (b) are associated directly
  with UV divergences in region (a) for the contributions from the gluon as well as
  from the quark hemisphere prescriptions. Since the two 
  prescriptions give identical results in region (a), they also have
  identical IR divergences in region (b).} The final outcome reads
\begin{align}\label{eq:S_cum}
 \Delta \mathcal{S}(K_L,K_R)=& \, \frac{\alpha_s^2 C_F T_F}{16
   \pi^2}\left[2\mathcal{F}_{qq}\left(\frac{K_L}{K_R}\right)-\frac{88}{9}+\frac{104}{27}\pi^2\right. \nn\\
   &+\left.\frac{32}{3}\zeta(3)\right] 
\end{align}
with the function
\begin{align}
  \mathcal{F}_{qq}(z)=& \, 16\,\Li_3(-z)-\frac{16}{3}\,\ln\,z \,
  \Li_2(-z)+\frac{8\pi^2}{9}\ln\, z \nn \\
  &-\frac{4(1-z)}{3(1+z)}\,\ln \,z \, , 
\end{align}
which is symmetric with respect to $z \leftrightarrow 1/z$. 

The result given in
Eq.~(\ref{eq:S_cum}) can be written as an integration over distributions in the
variables $k_l$ and $k_r$. Away
from the origin we should reproduce the structure in
Eq.~(\ref{eq:Dsoftmzerov1}) with (\ref{eq:f_qq_ml}) and (\ref{eq:f_g_ml}). We further
anticipate that the structure $[\frac{1}{k_l}]_{+} [\frac{1}{k_r}]_{+} f_{qq}(k_l/k_r)$ is part
of the final answer and compute its contribution to the cumulant using the fact
that $f_{qq}(0)=0$: ($\bar{K}_L=K_L/\mu$ etc.)  
\begin{align}\label{eq:cumulantplusdistr}
&\int^{\bar{K}_R}_0 d \bar{k}_r
\left[\frac{1}{\bar{k}_r}\right]_+\int^{\bar{K}_L}_0 d\bar{k}_l
\left[\frac{1}{\bar{k}_l}\right]_+
f_{qq}\left(\frac{\bar{k}_l}{\bar{k}_r}\right)\nn \\
&=\int^{\bar{K}_R}_0 d\bar{k}_r
\left[\frac{1}{\bar{k}_r}\right]_+ \int^{\frac{\bar{K}_L}{\bar{k}_r}}_0 \frac{
  f_{qq}(z)}{z}\,dz\nn\\ 
&=\int^{\frac{K_R}{K_L}}_0 dx \frac{1}{x}\left( \int^{1}_0 dz\,\frac{ f_{qq}(z/x)}{z}-C_{qq}\right)+C_{qq} \ln \, \bar{K}_R\nn\\
&\equiv\mathcal{F}_{qq}\left(\frac{K_L}{K_R}\right)+\frac{1}{2}C_{qq}\left(\ln
  \, \bar{K}_R+\ln\,\bar{K}_L\right) \,,
\end{align}
where  
\begin{align}\label{eq:C_qq}
C_{qq} \equiv \int^{\infty}_0 \frac{ f_{qq}(z)}{z}\,dz = -\frac{8}{3}+\frac{16\pi^2}{9}=C_g \, .
\end{align}
Note that the calculation is unambiguous owing to the property
$f_{qq}(0)=f_{qq}(\infty)=0$. Moreover, the order of the integrations can be exchanged, since
$f_{qq}(z)=f_{qq}(1/z)$. The result of Eq.~(\ref{eq:C_qq}) allows the
additional logarithms in the final equality of Eq.~(\ref{eq:cumulantplusdistr})
to be recombined with $C_g$ into distributions that yield the
correct cumulant in 
Eq.~(\ref{eq:S_cum}) compatible with Eq.~(\ref{eq:Dsoftmzerov1}). This gives the desired distributive expression for the
phase space misalignment 
correction:
\begin{align}
 &\mu^2 \Delta S(k_l,k_r,\mu)=\frac{\alpha_s^2 C_F T_F}{16 \pi^2}\left\{\delta(\bar{k}_l)\delta(\bar{k}_r)\left[-\frac{44}{9}+\frac{52}{27}\pi^2 \right.\right.\nn\\
 & + \left.\frac{16}{3}\zeta(3)\right]+\delta(\bar{k}_l)\left[\frac{\theta(\bar{k}_r)}{\bar{k}_r}\right]_{+}\left[\frac{8}{3}-\frac{16}{9}\pi^2\right] \nn \\
 &\left. +\left[\frac{\theta(\bar{k}_l)}{\bar{k}_l}\right]_{+}\left[\frac{\theta(\bar{k}_r)}{\bar{k}_r}\right]_{+} f_{qq}\left(\frac{k_l}{k_r}\right)+(k_l \leftrightarrow k_r) \right\} \, .
\end{align}
with the term $f_{qq}$ given in Eq.~(\ref{eq:f_qq_ml}). Combining this result with
Eq.~(\ref{eq:soft_gluon_massless}) we obtain the entire distributive structure of
the ${\cal O}(\alpha_s^2 C_F T_F n_f)$  massless quark corrections to the
momentum-space double hemisphere 
soft function valid for all $k_l,k_r\geq 0$,  
\begin{align}\label{eq:Snf_dist}
 &\mu^2 S_{n_f}(k_l,k_r,\mu)=\mu^2 Z_{S,n_f}(k_l,k_r,\mu)+\frac{\alpha_s^2 C_F T_F n_f}{16 \pi^2} \nn \\
 & \times \left\{\delta(\bar{k}_l)\,\delta(\bar{k}_r)\left[-\frac{68}{81}+\frac{37}{27}\pi^2+ \frac{28}{9}\zeta(3)\right]\right.  \nn \\
 & +\delta(\bar{k}_l)\left[\frac{\theta(\bar{k}_r)}{\bar{k}_r}\right]_{+} \left[-\frac{152}{27}- \frac{8\pi^2}{9}\right] \nn \\
  & + \delta(\bar{k}_l)\left[\frac{\theta(\bar{k}_r)\ln\,\bar{k}_r}{\bar{k}_r}\right]_{+}\frac{160}{9}- \delta(\bar{k}_l)\left[\frac{\theta(\bar{k}_r)\ln^2\bar{k}_r}{\bar{k}_r}\right]_{+}\frac{32}{3} \nn \\
  & + \left. \left[\frac{\theta(\bar{k}_l)}{\bar{k}_l}\right]_{+} \left[\frac{\theta(\bar{k}_r)}{\bar{k}_r}\right]_{+}f_{qq}\left(\frac{k_l}{k_r}\right) +(k_l \leftrightarrow k_r)\right\} \, ,
\end{align}
where we have included the massless flavor number $n_f$. The UV divergences which are absorbed into the soft function renormalization factor read
\begin{align}
  Z_{S,n_f}(k_l,k_r,\mu)= n_f Z_{S,m}(k_l,k_r,\mu) 
\end{align}
with $Z_{S,m}(k_l,k_r,\mu)$ given in Eq.~(\ref{eq:Z_Sm}). As already mentioned in Sec.~\ref{sec:Shemicomputation}, this result agrees for $k_l,k_r>0$ with
the naive $\epsilon\to 0$ expansion of the corresponding $d\neq 4$ result given
in Eq.~(31) of Ref.~\cite{Kelley:2011ng}. 

Interestingly, from the result
in Eq.~(\ref{eq:Snf_dist}) we can now also establish unambiguous rules for the
$\epsilon\to 0$ limit of the $d\neq 4$ results for the ${\cal O}(\alpha_s^2)$ corrections to
the momentum space double hemisphere discussed in
Refs.~\cite{Kelley:2011ng,Hornig:2011iu} valid for all $k_l,k_r\geq 0$: Writing
$f_{n_f}(z,\epsilon)= f_{n_f}^{(0)}(z)+\epsilon f_{n_f}^{(1)}(z)+\dots$ for the
opposite hemisphere correction given in Eqs.~(28),~(A4) and~(A5) of Ref.~\cite{Kelley:2011ng}, the
dictionary from their $d$-dimensional expression resulting
from Eq.~(\ref{eq:Snf_dist}) reads
\begin{align}\label{eq:replacement_rule}
 &\frac{\mu^{4\epsilon}}{(k_l
   k_r)^{1+2\epsilon}}f_{n_f}\left(\frac{k_l}{k_r},\epsilon\right) \nn \\
 & \stackrel{\epsilon\to 0}{\longrightarrow} \,\delta(\bar{k}_l)
 \,\delta(\bar{k}_r)\left[-\frac{1}{4\epsilon} \int^{\infty}_0 \frac{dz}{z}
  \left(f_{n_f}^{(0)}(z)+\epsilon
   f_{n_f}^{(1)}(z)\right)\right]\nn \\
   & + \left[\frac{\theta(\bar{k}_l)}{\bar{k}_l}\right]_{+}\left[\frac{\theta(\bar{k}_r)}{\bar{k}_r}\right]_{+}f_{n_f}^{(0)}\left(\frac{k_l}{k_r}\right)+\mathcal{O}(\epsilon)
\end{align}
with $f_{n_f}^{(0)}(z)=2f_{qq}(z)$, see Eq.~(\ref{eq:f_qq_ml}). Note that due to
the symmetry $k_l\leftrightarrow k_r$  ($f^{(n)}_{n_f}(z)=f^{(n)}_{n_f}(1/z)$)
and $f_{n_f}^{(n)}(0)=f_{n_f}^{(n)}(\infty)=0$ (for $n=0,1,2,...$) one has 
\begin{align}
 \int^{\infty}_0 dz \,\frac{\ln\,z}{z}\,  f^{(n)}_{n_f}(z)=0\, ,
\end{align}
so that corresponding contributions do not arise in Eq.~(\ref{eq:replacement_rule}).
The expansion given in Eq.~(\ref{eq:replacement_rule}) applies if
$f_{n_f}\left(0,\epsilon\right)=f_{n_f}\left(\infty,\epsilon\right)=0$. It is
straightforward to generalize to the case
$f\left(0,\epsilon\right)=f\left(\infty,\epsilon\right)\neq 0$. In this
situation we can define a new function
$\tilde{f}\left(z,\epsilon\right)=f\left(z,\epsilon\right)-f\left(0,\epsilon\right)$,
for which we can apply the rule in Eq.~(\ref{eq:replacement_rule}). For the
remaining constant $f\left(0,\epsilon\right)$ we can expand in $k_l$ and $k_r$
multiplicatively in the usual way generating products of additional single
variable distributions in $k_l$ and $k_r$.

\subsection{Result for the full ${\cal O}(\alpha_s^2)$  momentum-space double hemisphere soft function}

We are now in the position to write down the complete distributive expression for the
momentum-space double hemisphere soft function at ${\cal O}(\alpha_s^2)$
accounting for all gluonic, the massless as well as massive quark corrections. The
results for the pure gluonic corrections can be derived with the
help of the expansion rule (\ref{eq:replacement_rule}) and the results in
\cite{Kelley:2011ng,Hornig:2011iu}.  
The different structures for the unrenormalized soft function in the scheme, where the strong coupling evolves with
the $n_f$ massless and one massive flavor ($\alpha_s=\alpha_s^{(n_f+1)}$), read
\begin{align}
 S(k_l,k_r,\mu)= & \, S_{C_F}(k_l,k_r,\mu)+S_{C_A}(k_l,k_r,\mu) \nn \\
 &+S_{n_f}(k_l,k_r,\mu)
+ S_{m}(k_l,k_r,m,\mu)  \, ,
\end{align}
where $S_{n_f}(k_l,k_r,\mu)$ is the massless quark correction given in
Eq.~(\ref{eq:Snf_dist}) and $S_{m}(k_l,k_r,m,\mu)$ are the massive quark corrections in Eq.~(\ref{eq:DeltaS_tot}). Using the results from
\cite{Kelley:2011ng} for the $C_F C_A$-part and applying
(\ref{eq:replacement_rule}) we obtain 
\begin{align}\label{eq:SCA_dist}
 &\mu^2 S_{C_A}(k_l,k_r,\mu)=\mu^2 Z_{S,C_A}(k_l,k_r,\mu)+\frac{\alpha_s^2 C_F C_A}{16 \pi^2}\nn \\
 &\times \left\{\delta(\bar{k}_l)\,\delta(\bar{k}_r)\left[-\frac{1016}{81}-\frac{335}{108}\pi^2-\frac{77}{9}\zeta(3)+\frac{26}{45}\pi^4 \right] \right. \nn \\
 & +\delta(\bar{k}_l)\left[\frac{\theta(\bar{k}_r)}{\bar{k}_r}\right]_{+}\left[\frac{772}{27}+\frac{22\pi^2}{9}-36\zeta(3)\right]\nn \\
  & -\delta(\bar{k}_l)\left[\frac{\theta(\bar{k}_r)\ln\,\bar{k}_r}{\bar{k}_r}\right]_{+}\frac{536}{9}+\delta(\bar{k}_l)\left[\frac{\theta(\bar{k}_r)\ln^2\bar{k}_r}{\bar{k}_r}\right]_{+}\frac{88}{3} \nn \\
 & +\left[\frac{\theta(\bar{k}_l)}{\bar{k}_l}\right]_{+} \left[\frac{\theta(\bar{k}_r)}{\bar{k}_r}\right]_{+}\left[\frac{4\pi^2}{3}+ f_{gg}\left(\frac{k_l}{k_r}\right)\right] \nn \\
 &+(k_l \leftrightarrow k_r)\bigg\}
\end{align}
with
\begin{align}
 & f_{gg}(z)=\, 4\,\ln^2(1+z)-4\,\ln(1+z)\,\ln\,z-\frac{44}{3}\,\ln(1+z) \nn \\
 &+\frac{4z(12+21z+11z^2)}{3(1+z)^3}\,\ln\,z+\frac{8z}{3(1+z)^2} \, 
\end{align}
satisfying $f_{gg}(z)=f_{gg}(1/z)$ and $f_{gg}(0)=0$. The UV divergent contribution which adds to the soft function renormalization constant reads 
\begin{align}\label{eq:ZCA_dist}
 &\mu^2 Z_{S,C_A}(k_l,k_r,\mu)=\frac{\alpha_s^2 C_F C_A}{16 \pi^2}\delta(\bar{k}_l)\left\{\delta(\bar{k}_r)\left[\frac{11}{2\epsilon^3}\right.\right.\nn \\
 &+\left.\frac{1}{\epsilon^2}\left(-\frac{67}{18}+\frac{\pi^2}{6}\right)+\frac{1}{\epsilon}\left(-\frac{202}{27}+\frac{11\pi^2}{36}+7\zeta(3)\right)\right] \nn \\
 &\left. +\,\left[\frac{\theta(\bar{k}_r)}{\bar{k}_r}\right]_{+}\left[-\frac{22}{3\epsilon^2}+\frac{1}{\epsilon}\left(\frac{134}{9}-\frac{2\pi^2}{3}\right)\right]\right\}+(k_l \leftrightarrow k_r) \, .
\end{align}
To obtain this result
we have rewritten the function $f_{C_A}(z,\epsilon)$ given in Eq.~(17) of
Ref.~\cite{Kelley:2011ng} as $f_{C_A}(z,\epsilon)=\tilde{f}_{C_A}(z,\epsilon)+f_{C_A}(0,\epsilon)$
and proceeded as described at the end of Sec.~\ref{sec:distributionmassless}
for the expansion in terms of distributions. We have defined $2 f_{gg}(z)\equiv f_{C_A}^{(0)}(z)-f_{C_A}^{(0)}(0)$
with $f_{C_A}^{(0)}(z)$ given in Eq.~(A1) of Ref.~\cite{Kelley:2011ng} and
$f_{C_A}^{(0)}(0)\equiv8\pi^2/3$.

The distributive structure of the remaining gluonic $C_F^2$-corrections is
already known completely as it can be obtained from the exponentiation of the
one-loop result in position space~\cite{Hoang:2008fs}, and its $k_l$- and
$k_r$-dependence factorizes without any subtleties. For completeness we also
give the result for the $\mathcal{O}(\alpha_s^2 C_F^2)$ corrections to the unrenormalized soft function, 
\begin{align}\label{eq:SCF_dist}
 & \mu^2 S_{C_F}(k_l,k_r,\mu)=\mu^2 Z_{S,C_F}(k_l,k_r,\mu)+\frac{\alpha_s^2 C_F^2}{16 \pi^2}\nn\\
 & \times \left\{-\delta(\bar{k}_l)\delta(\bar{k}_r)\frac{11\pi^4}{180} + \delta(\bar{k}_l)\left[\frac{\theta(\bar{k}_r)}{\bar{k}_r}\right]_{+}64\zeta(3) \right. \nn \\
  &-\delta(\bar{k}_l)\left[\frac{\theta(\bar{k}_r)\ln\,\bar{k}_r}{\bar{k}_r}\right]_{+}\frac{40\pi^2}{3}+\delta(\bar{k}_l)\left[\frac{\theta(\bar{k}_r)\ln^3\bar{k}_r}{\bar{k}_r}\right]_{+}32 \nn \\
  &+\left.\left[\frac{\theta(\bar{k}_l)\ln\, \bar{k}_l}{\bar{k}_l}\right]_{+}\left[\frac{\theta(\bar{k}_r)\ln\,\bar{k}_r}{\bar{k}_r}\right]_{+}32+  (k_l \leftrightarrow k_r)\right\} \, 
\end{align}
with 
\begin{align}\label{eq:ZCF_dist}
 &\mu^2 Z_{S,C_F}(k_l,k_r,\mu)=\frac{\alpha_s^2 C_F^2}{16 \pi^2}\left\{\delta(\bar{k}_l)\delta(\bar{k}_r)\left[\frac{4}{\epsilon^4}-\frac{2\pi^2}{\epsilon^2}\right.\right.\nn\\
 &-\left.\frac{32\zeta(3)}{\epsilon}\right]+\delta(\bar{k}_l)\left[\frac{\theta(\bar{k}_r)}{\bar{k}_r}\right]_{+}\left[-\frac{16}{\epsilon^3}+\frac{20\pi^2}{3\epsilon}\right] \nn \\
  & +\delta(\bar{k}_l)\left[\frac{\theta(\bar{k}_r)\ln\,\bar{k}_r}{\bar{k}_r}\right]_{+}\frac{48}{\epsilon^2}-\delta(\bar{k}_l)\left[\frac{\theta(\bar{k}_r)\ln^2\bar{k}_r}{\bar{k}_r}\right]_{+}\frac{48}{\epsilon} \nn \\
  & +\left[\frac{\theta(\bar{k}_l)}{\bar{k}_l}\right]_{+} \left[\frac{\theta(\bar{k}_r)}{\bar{k}_r}\right]_{+}\frac{8}{\epsilon^2} \nn \\
 &\left. -\left[\frac{\theta(\bar{k}_l)}{\bar{k}_l}\right]_{+}\left[\frac{\theta(\bar{k}_r)\ln\,\bar{k}_r}{\bar{k}_r}\right]_{+}\frac{32}{\epsilon}+(k_l \leftrightarrow k_r)\right\} 
\end{align}
for the UV divergent terms.
We have checked analytically that the cumulants generated by
Eqs.~(\ref{eq:Snf_dist}),~(\ref{eq:SCA_dist}),~(\ref{eq:SCF_dist}) agree with the
expressions given in Eqs.~(50) and (53) of Ref.~\cite{Kelley:2011ng} and Eqs.~(3.36)-(3.43) of Ref.~\cite{Hornig:2011iu}.\footnote{Note that the constants $s_2^{[n_f]}$ and $s_2^{[C_F C_A]}$ in Eq.~(3.42) of Ref.~\cite{Hornig:2011iu} should be converted from position space to momentum space by including the terms listed in Eq.~(45) of Ref.~\cite{Kelley:2011ng}.}
Furthermore, we have found agreement to the resulting expressions for the corresponding ${\cal
  O}(\alpha_s^2)$ corrections to the thrust soft
function given in Eq.~(41) of Ref.~\cite{Kelley:2011ng} (see also \cite{Monni:2011gb}), and furthermore to the heavy jet mass constant in Eq.~(42) of Ref.~\cite{Kelley:2011ng}.
Moreover, the position space representation of the massless quark and gluon corrections, i.e. the Fourier
transformations of Eqs.~(\ref{eq:Snf_dist}),~(\ref{eq:SCA_dist}),~(\ref{eq:SCF_dist}),
agree with the ones given in Eqs.~(3.30)-(3.35) of Ref.~\cite{Hornig:2011iu}. 

\section{Projection onto thrust}\label{sec:projections}

From the massive quark corrections to the ${\cal O}(\alpha_s^2)$ double
hemisphere soft function we can derive the corresponding soft function
corrections for a few other event shape variables. Here we investigate the most prominent projection, namely thrust. We define the thrust variable by\footnote{ 
Note that $\tau$ is normalized with the c.m.\ energy $Q$,
which is the sum of all energies and also agrees with the variable
2-jettiness~\cite{Stewart:2010tn}. For massless decay products this agrees with
the common definition, which is normalized to the sum of momenta $\sum_i
|\vec{p}_i|$.} 
\begin{align}
\tau=1-T \,=\, 1-\sum_i \frac{ |\vec{n} \cdot
\vec{p}_i|}{\sum_j |E_j|}
\,=\, 1-\sum_i \frac{ |\vec{n} \cdot
\vec{p}_i|}{Q}\,,
\end{align}
where $\vec{n}$ is the thrust axis and the sum is performed over all final state
particles with momenta $\vec{p}_i$ and energies $E_i$. 
The partonic soft function for the thrust distribution can be easily obtained from the linear relation 
\begin{align}
 \tau = \frac{M_l^2 +M_r^2}{Q^2} +\mathcal{O}\left(\frac{M_{l,r}^4}{Q^4}\right)
\end{align}
in the dijet limit, which yields
\begin{align}\label{eq:soft_thrust}
  S_{\tau}(\ell,m,\mu)= \int dk_r \, dk_l \, \delta(\ell-k_r-k_l) \, S(k_r,k_l,m,\mu)  \, .
\end{align}
We can split the ${\cal O}(\alpha_s^2 C_F T_F)$ massive quark corrections to the
partonic thrust soft function into
\begin{align}
 S_{\tau,m}(\ell,m,\mu)=& \, Z_{S,\tau}(\ell,\mu)+ S^{(g)}_{\tau,\rm virt}(\ell,m,\mu)  \nn \\
   &+ S^{(g)}_{\tau,\rm
   real}(\ell,m,\mu) + \Delta S_{\tau}(\ell,m) \, ,
\label{softthrust1}
\end{align}
according to Eqs.~(\ref{eq:soft_gluon_massive}) and (\ref{eq:DeltaS_tot}). In Eq.~(\ref{softthrust1}) the term
 $S^{(g)}_{\tau,\rm virt}$ ($S^{(g)}_{\tau,\rm real}$) corresponds to the virtual (real) massive quark radiation piece
coming from the gluon hemisphere prescription, while $\Delta S_{\tau}(\ell,m)$ is the finite phase space
misalignment correction due to the physical quark hemisphere prescription.
These are related to the corresponding double hemisphere results of
Eqs.~(\ref{eq:soft_gluon_massive_virt}),~(\ref{eq:soft_gluon_massive_real})
and~(\ref{eq:Dsoftdoublenum}). 
For the gluon hemisphere contributions the convolution according to
Eq.~(\ref{eq:soft_thrust}) is straightforward and we obtain  ($\bar \ell = \ell/\mu$)
\begin{align}
&\mu\, S^{(g)}_{\tau,\rm virt}(\ell,m,\mu)= \frac{\alpha_s^2 C_F T_F}{16\pi^2}  \left\{\delta(\bar{\ell})\left[-\frac{8}{9}L_m^3-\frac{40}{9}L_m^2 \right.\right. \nn \\
& +\left. \left(-\frac{448}{27}+\frac{8\pi^2}{9}\right)L_m -\frac{656}{27}+\frac{10\pi^2}{27}+\frac{56}{9}\zeta(3) \right] \nn \\
 &+  \left[\frac{\theta(\bar{\ell})}{\bar{\ell}}\right]_{+}\left[\frac{16}{3}L_m^2+\frac{160}{9}L_m+\frac{448}{27}\right] \nn \\
 &- \left.\left[\frac{\theta(\bar{\ell})\ln
\, \bar{\ell}}{\bar{\ell}}\right]_{+} \frac{64}{3}L_m \right\}  \, ,  \label{eq:Sg_thrustvirt}\\ 
& \mu\, S^{(g)}_{\tau,\rm real}(\ell,m,\mu)= \frac{\alpha_s^2 C_F T_F}{16\pi^2}
 \, \theta(\ell-2m)\,\frac{2}{\bar{\ell}}
R\left(\sqrt{1-\frac{4m^2}{\ell^2}}\right) \, ,
\label{eq:Sg_thrustreal}
\end{align} 
where the function $R\left(w\right)$ is defined in Eq.~(\ref{eq:Rfunction}). The UV divergences that contribute to the soft function renormalization constant read 
\begin{align}
 & \mu\, Z_{S,\tau}(\ell,\mu)= \frac{\alpha_s^2 C_F T_F}{16\pi^2} \left\{\delta(\bar{\ell})\left[-\frac{4}{\epsilon^3}+\frac{20}{9\epsilon^2} \right.\right.\nn \\
 &\left.+\left.\frac{1}{\epsilon}\left(\frac{112}{27}-\frac{2\pi^2}{9}\right)\right] + \left[\frac{\theta(\bar{\ell})}{\bar{\ell}}\right]_{+} \left[\frac{16}{3\epsilon^2}-\frac{80}{9\epsilon}\right]\right\} \, .
\end{align}

\begin{figure}
  \begin{center}
  \subfigure{\epsfig{file=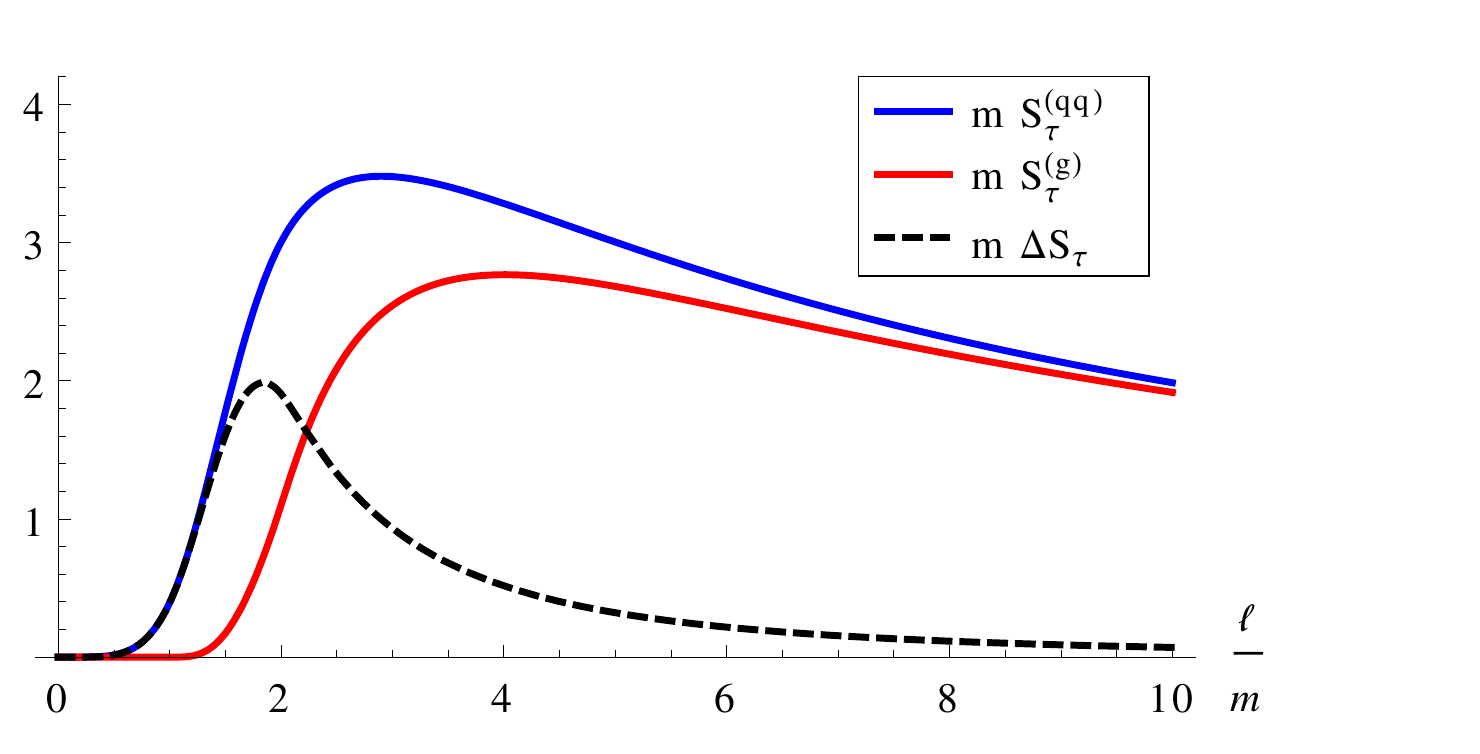,width=1.1\linewidth,clip=}}
  \caption{The contributions to the thrust soft function from the opposite hemisphere phase space, for the quark hemisphere prescription $m\,S^{(qq)}(\ell,m)$ (blue), gluon hemisphere prescription $m\,S^{(g)}(\ell,m)$ (red) and the difference giving $m\,\Delta S_{\tau} (\ell,m)$ (black, dashed).}
  \label{fig:soft_thrust}
  \end{center}
\end{figure}

The contribution from the phase space misalignment correction $\Delta S_{\tau}(\ell,m)$ can
be written analogously to Eq.~(\ref{eq:DeltaS_hemiPS}),
\begin{align}\label{eq:DeltaS_thrustPS}
& \Delta S_{\tau} (\ell,m)= \frac{\alpha_s^2 C_F T_F}{16\pi^2}\int dq^{-} \int
 dk^{+} \int dq^{+} \int dk^{-} \nn \\
& \times \theta(k^{-}-k^{+})\,\theta(q^{+}-q^{-})\,\theta(k^{+}k^{-}-m^2)\,\theta(q^{+}q^{-}-m^2) \nn \\ 
 &\times\,\theta(k^{-}+k^{+})\,\theta(q^{+}+q^{-}) \left[\delta(\ell-k^{+}-q^{-}) \right. \nn \\
 &-\theta(k^{-}+q^{-}-k^{+}-q^{+})\,\delta(\ell-k^{+}-q^{+})\nn \\
 & -\left.\theta(k^{+}+q^{+}-k^{-}-q^{-})\,\delta(\ell-k^{-}-q^{-})\right] \nn \\
 & \times f_m(k^{+},k^{-},q^{+},q^{-},m)
\end{align}
where $f_m$ has been given in Eq.~(\ref{eq:fm}). 
It can be calculated
numerically using the Cuba library \cite{Hahn:2004fe}. For large values of $\ell/m$ $\Delta
S_{\tau}(\ell,m)$ involves strong cancellations in the difference between its
quark and gluon hemisphere contributions (see also Eq.~(\ref{eq:phase-space-diff})), which can lead to numerical instabilities.
This is illustrated in Fig.~\ref{fig:soft_thrust} where $\Delta
S_{\tau}(\ell,m)$ is shown together with its two contributions from both
prescriptions. An alternative way to
compute $\Delta S_{\tau}(\ell,m)$ for large values of $\ell/m$ can be achieved by evaluating the cumulant,
where the quark and gluon hemisphere contributions can be combined prior to integration, and by differentiating numerically afterwards (see the appendix). 

In the massless limit $\Delta S_\tau$ becomes a delta function and gives ($\bar{\ell}=\ell/\mu$)
\begin{align}\label{eq:DeltaS_massless}
  \mu \,\Delta S_{\tau}(\ell,m)  \stackrel{m \rightarrow 0}{\longrightarrow}  \frac{\alpha_s^2 C_F T_F}{16\pi^2} \delta(\bar{\ell})\left\{ -\frac{64}{9}+\frac{104\pi^2}{27}-\frac{64\zeta(3)}{3} \right\} \, .
\end{align}
Together with the massless limits of Eqs.~(\ref{eq:Sg_thrustvirt}),~(\ref{eq:Sg_thrustreal}) this yields
\begin{align}
& \mu \, S_{\tau,m}(\ell,m\rightarrow 0,\mu)= \mu \, Z_{S,\tau}(\ell,\mu) + \frac{\alpha_s^2 C_F T_F}{16\pi^2} \nn\\
& \times \left\{\delta(\bar{\ell}) \left[\frac{80}{81}+\frac{74\pi^2}{27}-\frac{232}{9}\zeta(3) \right]+\left[\frac{\theta(\bar{\ell})}{\bar{\ell}}\right]_{+}\left[-\frac{448}{27}\right. \right. \nn \\ 
 & \left.+ \left.\frac{16\pi^2}{9}\right] +\left[\frac{\theta(\bar{\ell})\,\ln \,
       \bar{\ell}}{\bar{\ell}}\right]_{+}\frac{320}{9} -
   \left[\frac{\theta(\bar{\ell})\,\ln^2
       \bar{\ell}}{\bar{\ell}}\right]_{+}\frac{64}{3} \right\} \,,
\end{align}
which is the result known for one massless quark flavor \cite{Kelley:2011ng,Monni:2011gb}.

\begin{figure}
  \begin{center}
  \subfigure{\epsfig{file=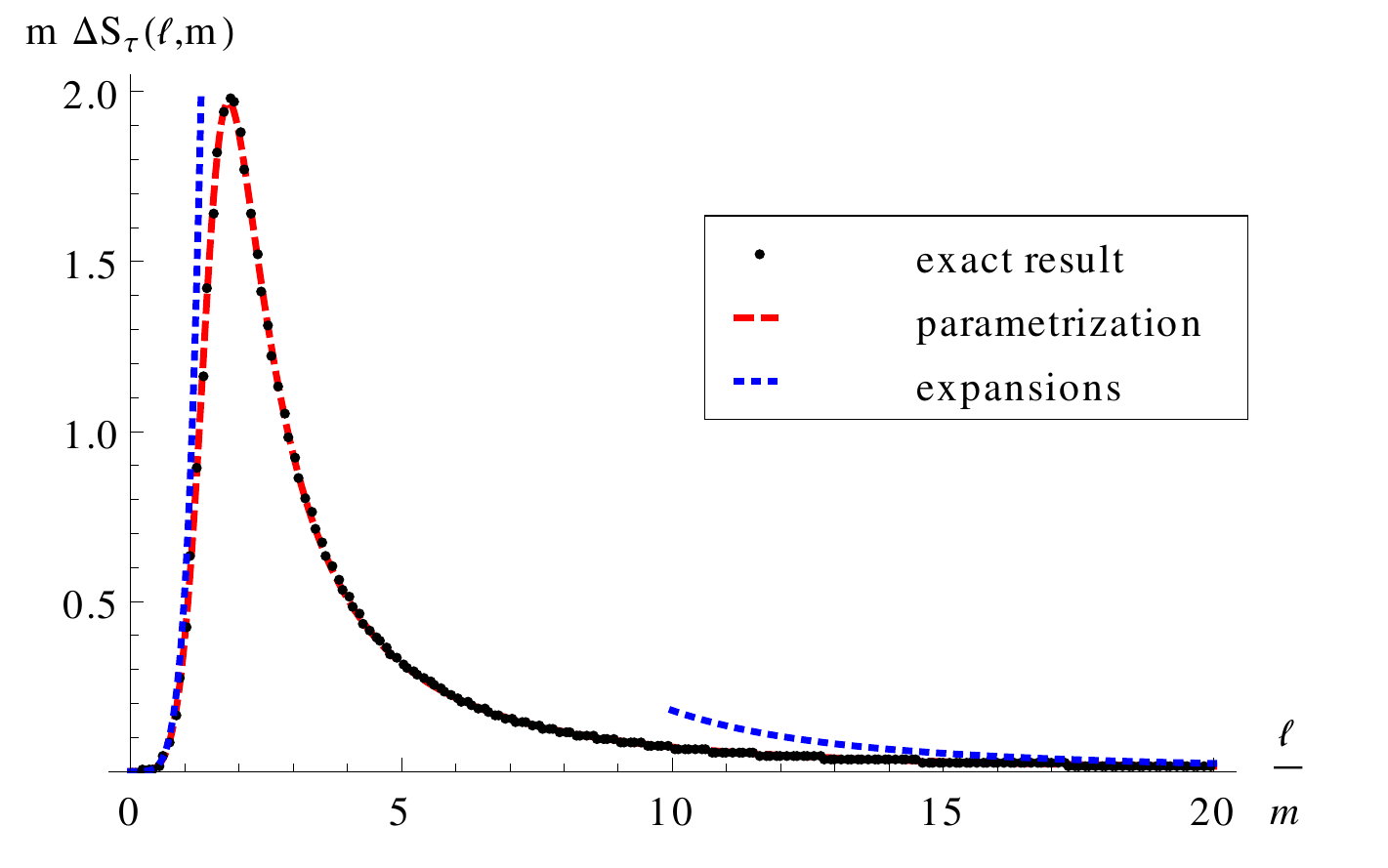,width=\linewidth,clip=}}
  \subfigure{\epsfig{file=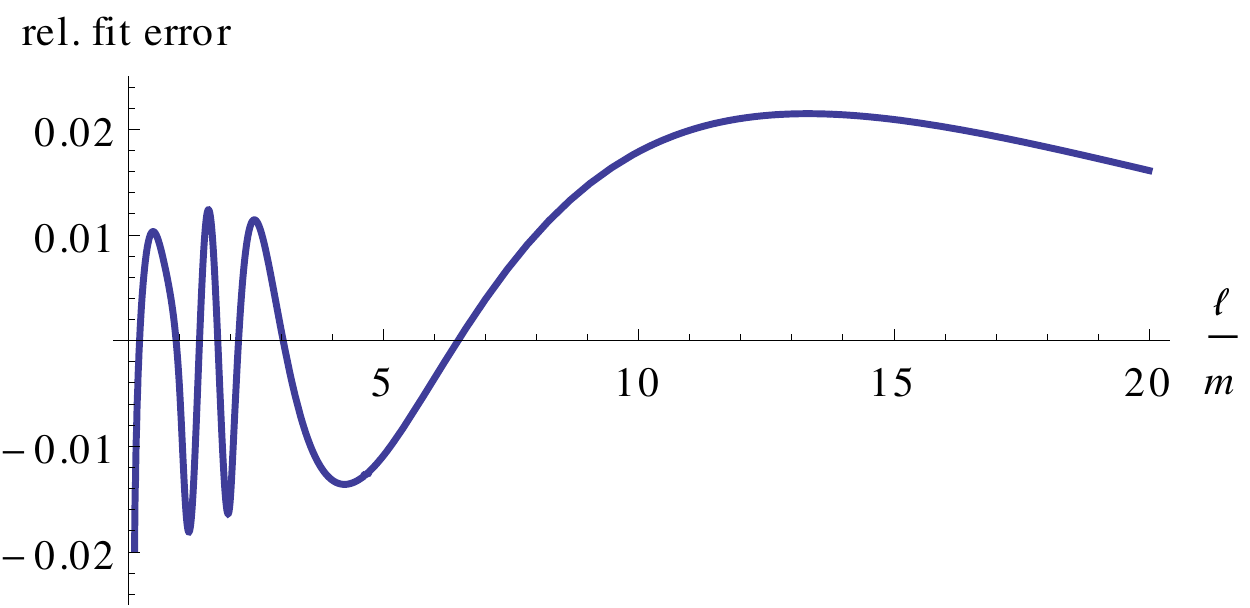,width=\linewidth,clip=}} 
  \caption{Above: Phase space misalignment correction $m\, \Delta S_{\tau}(\ell,m)$ (black dots) together with the fit of Eq.~(\ref{eq:DeltaS_parametrization}) (red,dashed) and its asymptotic expansions (blue,dotted), normalized by the prefactor $\alpha_s^2 C_F T_F/16\pi^2$. The fit is almost indistinguishable from the exact function. Below: The relative difference between fit and exact function for $\Delta S_{\tau}(\ell,m)$ (compared to an interpolation).}
  \label{soft_func_fit}
  \end{center}
\end{figure}
 
We aim at providing a parametrization of $\Delta S_{\tau}(\ell,m)$, which can be
used for a numerical analysis of mass effects for the thrust distribution. For
this purpose, we perform asymptotic expansions for small and large ratios
$\ell/m$. First we consider the expansion for small thrust momenta or large quark
masses, which can be obtained from integrating Eq.~(\ref{eq:f_qq_MM}), yielding 
\begin{align}\label{eq:DeltaS_highmass}
 \mu \, \Delta S_\tau (\ell,m)  \stackrel{\ell \ll m}{\longrightarrow}
 \frac{\alpha_s^2 C_F T_F}{16\pi^2} \,\frac{1}{\bar{\ell}} \frac{8
   \ell^6}{15 m^6}  \left[1+\mathcal{O}\left(\frac{\ell^2}{m^2}\right)\right] \,
 . 
\end{align}
Note that there is no threshold, below which this contribution vanishes. The
expansion for large thrust momenta is more challenging and
described in the appendix. The final result reads 
\begin{align}\label{eq:DeltaS_lowmass}
 &\mu\, \Delta S_\tau (\ell,m)  \stackrel{\ell \gg m}{\longrightarrow}
 \frac{\alpha_s^2 C_F T_F}{16\pi^2} \frac{1}{\bar{\ell}} \frac{m^2}{\ell^2}
 \left[8  \ln^2\left(\frac{m^2}{\ell^2}\right) \right.\nn \\
 &+ \left. 80 \ln\left(\frac{m^2}{\ell^2}\right)+\frac{640}{3}+\frac{292
     \pi^2}{45}+32\pi\right]\left[1+\mathcal{O}\left(\frac{m}{\ell}\right)\right] \, . 
\end{align}
A possible parametrization of $\Delta S_\tau (\ell)$ can be given by a
Pad\'{e}-type rational function multiplying some logarithmic terms. We adopt an
analytic ansatz that is capable of yielding the asymptotic behaviors of Eqs.~(\ref{eq:DeltaS_highmass}) and ~(\ref{eq:DeltaS_lowmass}) and
has a finite normalization, 
\begin{align}\label{eq:DeltaS_parametrization}
 & \Delta S_\tau (x=\ell/m)\bigg|_{\rm fit} =  \frac{\alpha_s^2 C_F T_F}{16\pi^2} \,  \frac{1}{m} \nn \\
 & \times\frac{x^5\left(a \, \ln^2\left(1+x^2\right)+b \, \ln\left(1+x^2\right) + c\right)}{d x^8 +e x^7 +f x^6 +g x^4 + h x^3 + j x^2 +1} \, . 
\end{align}
with $a=8d$, $b=-80d$, $c=8/15$ and $d=6/(2400+360\pi+73\pi^2)$ fixed by requiring the correct asymptotic behavior. The remaining 5 parameters were obtained using a $\chi^2$-fit with the constraint of satisfying the correct normalization corresponding to the massless analytic limit given in Eq.~(\ref{eq:DeltaS_massless}). We get 
\begin{align}
 &  e = 0.0117 \,, \, f = 0.100 \, , \, g = -0.502 \, , \nn \\
 & h = 0.747 \, , \,  j = -0.180 \, .
\end{align}
These fitted values correspond to a local minimum of the
$\chi^2$-function which has the feature that the relative error of the fit function
to the exact function does not exceed $3\%$ anywhere, and is around $1\%$ in the peak region, where the bulk of the
contribution arises. The exact and fitted results together with the asymptotic expansions
are displayed in Fig. \ref{soft_func_fit}.

\begin{figure}
\centering
\subfigure{\includegraphics[width=\linewidth]{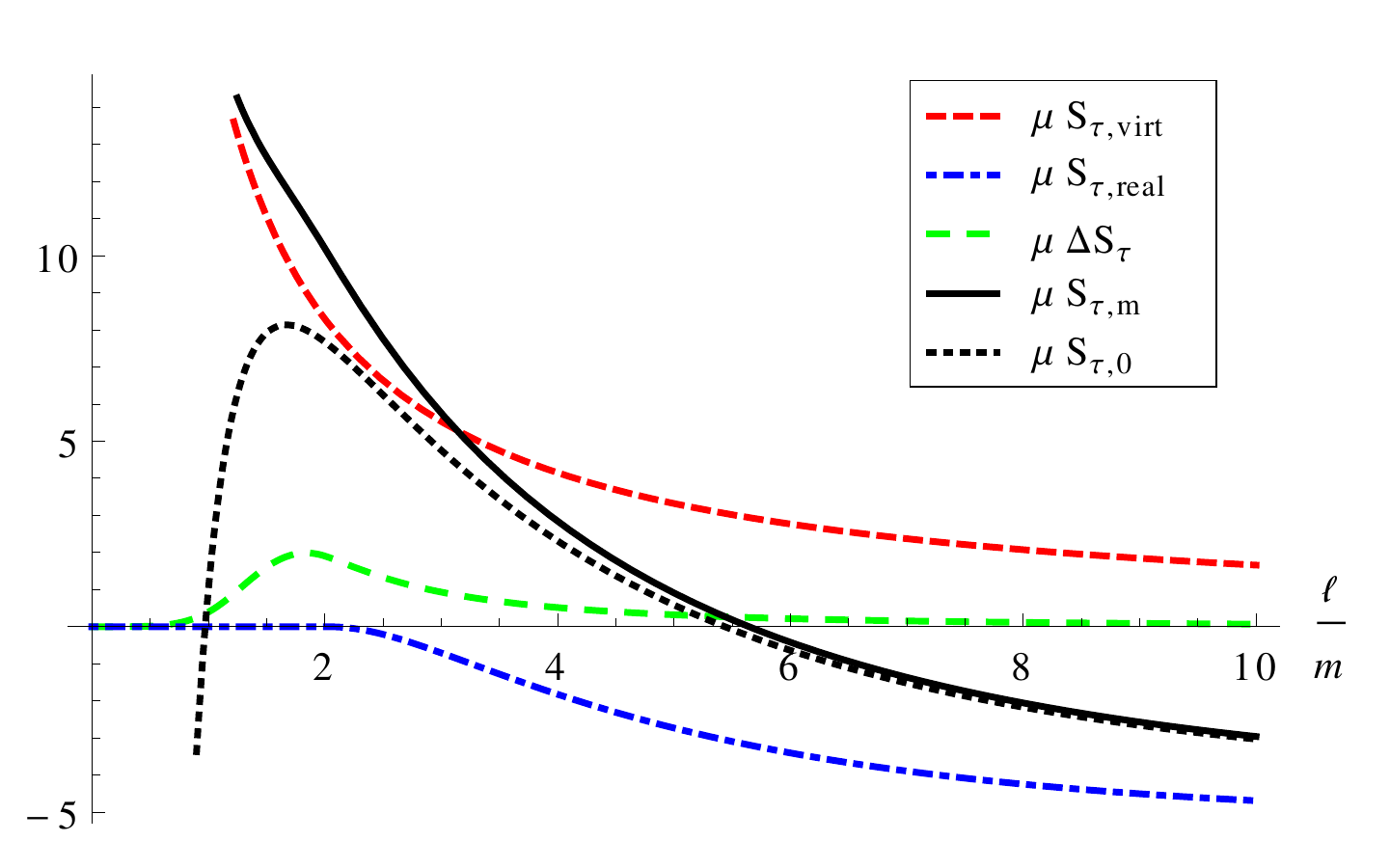}}
\subfigure{\includegraphics[width=\linewidth]{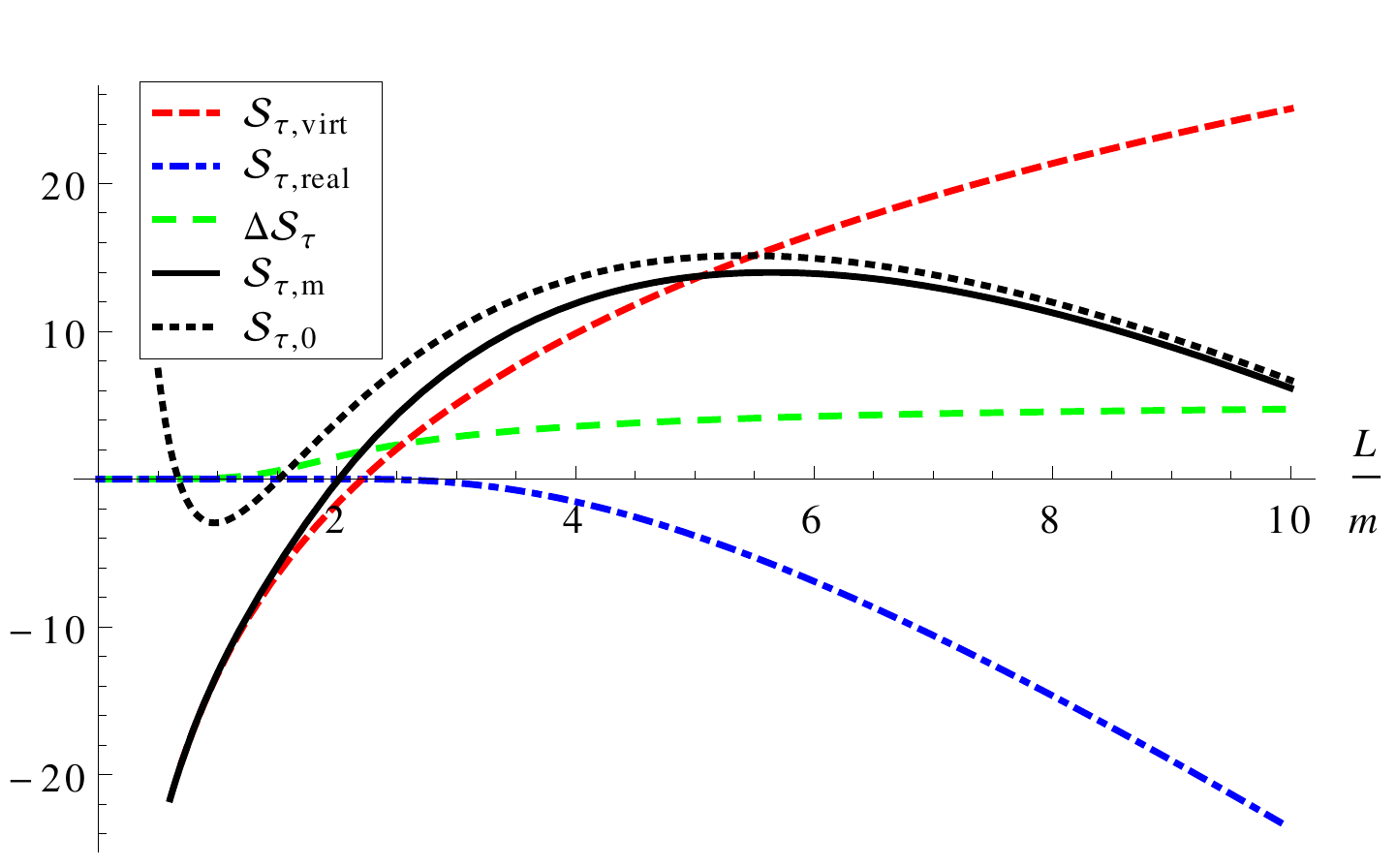} }
\caption{Plots of the massive quark contributions to the thrust soft function (upper panel) and its cumulant (lower panel) for $\mu=m$, normalized by the prefactor $\alpha_s^2 C_F T_F/16\pi^2$. $S_{\tau,{\rm virt}}$, $S_{\tau,{\rm real}}$ and $\Delta S_{\tau}$ denote the virtual, real radiation and phase space misalignment contributions described in the text, $S_{\tau,m}$ is their sum, i.e. the full massive contribution, and $S_{\tau,0}$ is the massless contribution. The corresponding descriptions also hold for the cumulant terms.}
\label{fig:softfct_plots}
\end{figure}

The three components of the $\mathcal{O}(\alpha_s^2 C_F T_F)$ massive quark corrections to the renormalized thrust soft function, their sum and the massless limit are displayed together with the cumulants in Fig.~\ref{fig:softfct_plots} for $\mu=m$ as a function of $\ell/m$ and $L/m$, respectively. We see that the phase space misalignment correction represents a relatively small contribution.

\section{Renormalon subtractions with secondary massive
particles}\label{sec:renormalon_subtraction}
 
The complete soft function is a convolution of
the partonic soft function, describing perturbative corrections at the
soft scale,
and the nonperturbative hadronic soft function~\cite{Hoang:2007vb}. Using
dimensional
regularization and the $\MS$ scheme for UV divergences entails that the
interface
between perturbative and nonperturbative contributions suffers from IR
renormalon
problems. These are related to contributions from very
small
momenta entering in the perturbative computations and lead to factorially
enhanced
coefficients of the high-order perturbation series which can render the
determination of
the nonperturbative parameters in the hadronic soft function unstable. In the
gap formalism for the soft function~\cite{Hoang:2007vb,Hoang:2008fs} one can eliminate
the renormalon problem
for the leading ${\cal O}(\Lambda_{\rm QCD})$ power correction that arises in
the operator production expansion (OPE) of the
soft function for $k_l\sim k_r\gg \Lambda_{\rm QCD}$. This is achieved through
a perturbative subtraction that eliminates order-by-order the leading
power IR
sensitivity of the partonic soft function.
The name of the gap formalism arises from the fact that the subtraction is
physically
related to the minimal hadronic energy deposit $\Delta\sim\Lambda_{\rm
QCD}$ in
the two hemispheres and can thus be implemented through a shift in the
momentum
arguments $k_l$ and $k_r$ of the partonic soft function. So the subtracted
partonic soft function, which is free of the  ${\cal O}(\Lambda_{\rm QCD})$
renormalon has the form~\cite{Hoang:2007vb}
\begin{align}\label{Spartsubtracted}
S_{\rm part}(k_l-\delta(R,\mu),k_r-\delta(R,\mu),\mu)\,,
\end{align}
where $\delta(R,\mu)$ is a properly defined perturbative series. A
very convenient definition is \cite{Hoang:2008fs}
\begin{align}\label{eq:delta_hemi}
 &\delta(R,\mu)=  \frac{R e^{\gamma_E}}{2} \left(\frac{d}{d\,\ln(ix_l)} \right. \nn \\
 & + \left. \frac{d}{d\,\ln(ix_r)}\right) \left. \ln \,
\tilde{S}_{\rm{part}}(x_l,x_r,\mu)\right|_{x_l=x_r=(i R
e^{\gamma_E})^{-1}} ,
\end{align}
where $\tilde{S}$ is the configuration space partonic soft function
\begin{align}
  \tilde{S}_{\rm{part}}(x_l,x_r,\mu)=\int dk_l \, dk_r \,S_{\rm
part}(k_l,k_r,\mu) \, e^{-i k_l x_l}e^{-i k_r x_r} \, .
\end{align}
The subtracted soft function in expression~(\ref{Spartsubtracted}) must be
expanded out order-by-order in powers of the strong coupling, and the
definition
in Eq.~(\ref{eq:delta_hemi}) ensures that the subtraction has the correct
normalization and the proper behavior at low as well as in higher orders in
the perturbation series. Renormalon-free soft functions based on the gap
subtraction given in Eq.~(\ref{eq:delta_hemi}) for massless quarks have been
used in the event shape analyses~\cite{Abbate:2010xh,Abbate:2012jh}.
 
For the ${\cal O}(\alpha_s^2 C_F T_F)$ massive quark corrections to the
partonic
soft function the finite quark mass provides
an infrared cutoff for the virtuality of the exchanged gluon such that the
factorial growth of the coefficients at large orders in perturbation theory is suppressed and, in
principle,
a corresponding subtraction for the massive quark corrections appears
unnecessary.
However, implementing the gap scheme along the lines of
Eqs.~(\ref{Spartsubtracted}) and (\ref{eq:delta_hemi}) is useful also for
the ${\cal
  O}(\alpha_s^2 C_F T_F)$ massive quark corrections in order to have a
smooth interpolation of the gap scheme parameters to the massless quark
limit.
This is in analogy to using the $n_f+1$ dynamical flavor scheme for the
renormalization group evolution of the strong coupling for $n_f$ massless and
one massive quark flavors for renormalization scales just above the quark mass.
However, for the gap-subtracted soft function in Eq.~(\ref{Spartsubtracted})
the subtraction series $\delta$ will in general be mass-dependent since it
represents infrared-sensitive perturbative contributions.
 
We note that one can derive the gap subtraction also directly from the thrust
soft function since the Fourier space partonic soft function is related to
the
double hemisphere partonic soft function by the simple relation
$\tilde{S}_{\tau,\rm{part}}(x,\mu) = \tilde{S}_{\rm{part}}(x,x,\mu)$. So
we have
the identity
\begin{align}\label{eq:delta_thrust}
 \delta(R,\mu)= \left. \frac{R e^{\gamma_E}}{2} \frac{d}{d\,\ln(ix)}
   \left[\ln \, \tilde{S}_{\tau,\rm{part}}(x,\mu)\right] \right|_{x=(i R
   e^{\gamma_E})^{-1}} \, ,
\end{align}
which we use to determine the gap subtraction in the following.
 
Following the form of Eq.~(\ref{softthrust1}) we parametrize the gap subtraction coming
from the $\mathcal{O}(\alpha_s^2 C_FT_F)$ massive quark corrections in the form
\begin{align}\label{eq:deltam}
& \delta_{m}(R,m,\mu)= \frac{\alpha_s^2(\mu) C_F T_F}{(4\pi)^2} R e^{\gamma_E} \nn \\
& \times \Big[\,
h_{\rm virt}(R,m,\mu)+h_{\rm real}(R,m)+h_{\Delta}(R,m)
\,\Big]\, ,
\end{align}
where the three terms in the brackets arise from the results for
$S^{(g)}_{\tau,\rm virt}$, $S^{(g)}_{\tau,\rm real}$ and $\Delta S_{\tau}$
given
in Eqs.~(\ref{eq:Sg_thrustvirt}), (\ref{eq:Sg_thrustreal}) and
(\ref{eq:DeltaS_thrustPS}). We emphasize that the results are given within
the scheme with $n_f+1$ dynamical quark flavors
($\alpha_s=\alpha_s^{(n_f+1)}$).
We obtain ($L_m=\ln(m^2/\mu^2)$)
\begin{align}\label{eq:delta_virt}
 h_{\rm virt}(R,m,\mu)=
-\,\frac{16}{3}L_m\ln\left(\frac{\mu^2}{R^2}\right)-\frac{8}{3}
L_m^2-\frac{80}{9}L_m-\frac{224}{27} \, ,
\end{align}
and after some lengthy analytical calculation ($z\equiv 2m/(R e^{\gamma_E})$)
\begin{align}
& h_{\rm real}(R,m)= \frac{16}{3} G^{3,0}_{1,3}\left(\begin{matrix} 1
\\0,0,0 \end{matrix}\middle\vert
\frac{z^2}{4}\right)-\frac{160}{9}K_0(z)\nn \\
& +z\bigg[\frac{160}{9}K_1(z)-8\pi\bigg]+
z^2\bigg[-\frac{16}{27}K_2(z) \nn \\
& + 8\pi(K_0(z)L_{-1}(z)+K_1(z)L_0(z))\bigg]+z^3\bigg[\frac{16}{27}K_1(z)-\frac{8}{27}\pi\bigg] \nn \\
 & + z^4\frac{8}{27}\pi\bigg[K_0(z)L_{-1}(z)+K_1(z)L_0(z)\bigg]\,,
\label{eq:delta_real}
 \end{align}
where $K_n$ are Bessel functions. $G^{m,n}_{p,q}$ and $L_n$ denote the less known Meijer G and Struve functions, for which some explicit integral representations are provided in appendix~\ref{app:b}.
The contribution from the phase space misalignment correction can again
not be
given in closed analytic form and, using Eqs.~(\ref{eq:DeltaS_hemiPS}),
(\ref{eq:soft_thrust}) and~(\ref{eq:delta_thrust}), reads 
\begin{align}\label{eq:DeltaS_thrust_ren}
 & h_{\Delta}(R,m)= -\frac{1}{2} %
(R e^{\gamma_E})^{-1}
\int dq^{-} \int dk^{+} \int dq^{+} \int dk^{-} \nn \\
& \times  \theta(k^{-}-k^{+})\,\theta(q^{+}-q^{-})\,\theta(k^{+}k^{-}-m^2)\,\theta(q^{+}q^{-}-m^2)
\nn \\
 &\times\theta(k^{+}+k^{-})\,\theta(q^{+}+q^{-})\left[(q^{-}+k^{+}) \,e^{-\frac{q^{-}+k^{+}}{R
e^{\gamma_E}}}\right.\nn \\
& \hspace{0.5cm} -\theta(k^{-}+q^{-}-k^{+}-q^{+})(k^{+}+q^{+}) \,
e^{-\frac{k^{+}+q^{+}}{R e^{\gamma_E}}}\nn \\
 & \hspace{0.5cm} -\left.\theta(k^{+}+q^{+}-k^{-}-q^{-})\,(k^{-}+q^{-})\,\,
   e^{-\frac{k^{-}+q^{-}}{R e^{\gamma_E}}}\right] \nn \\
& \times f_m(k^{+},k^{-},q^{+},q^{-},m) \, .
 \,
\end{align}
\begin{figure}
\centering
\subfigure{\includegraphics[width=\linewidth]{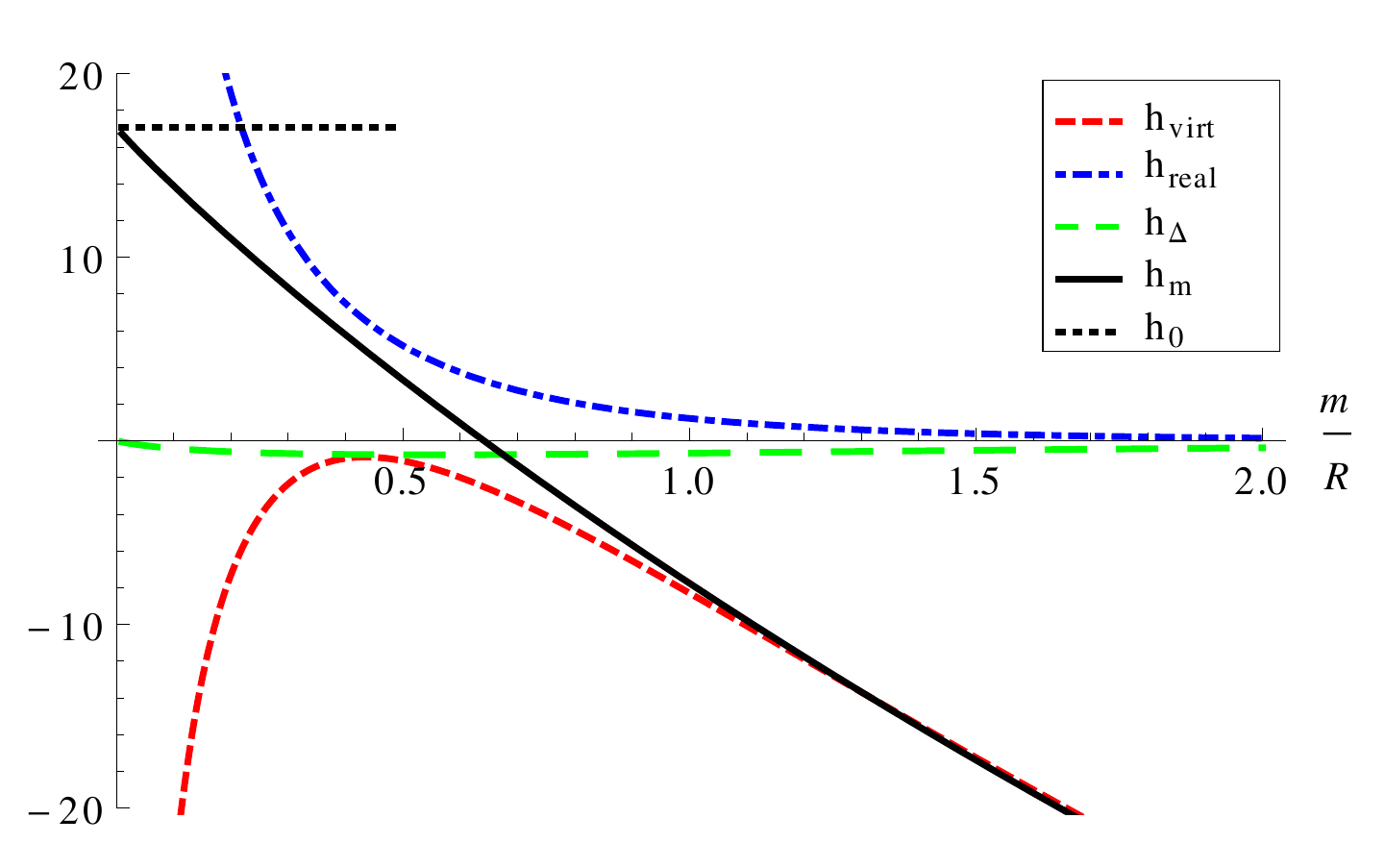}}
\subfigure{\includegraphics[width=\linewidth]{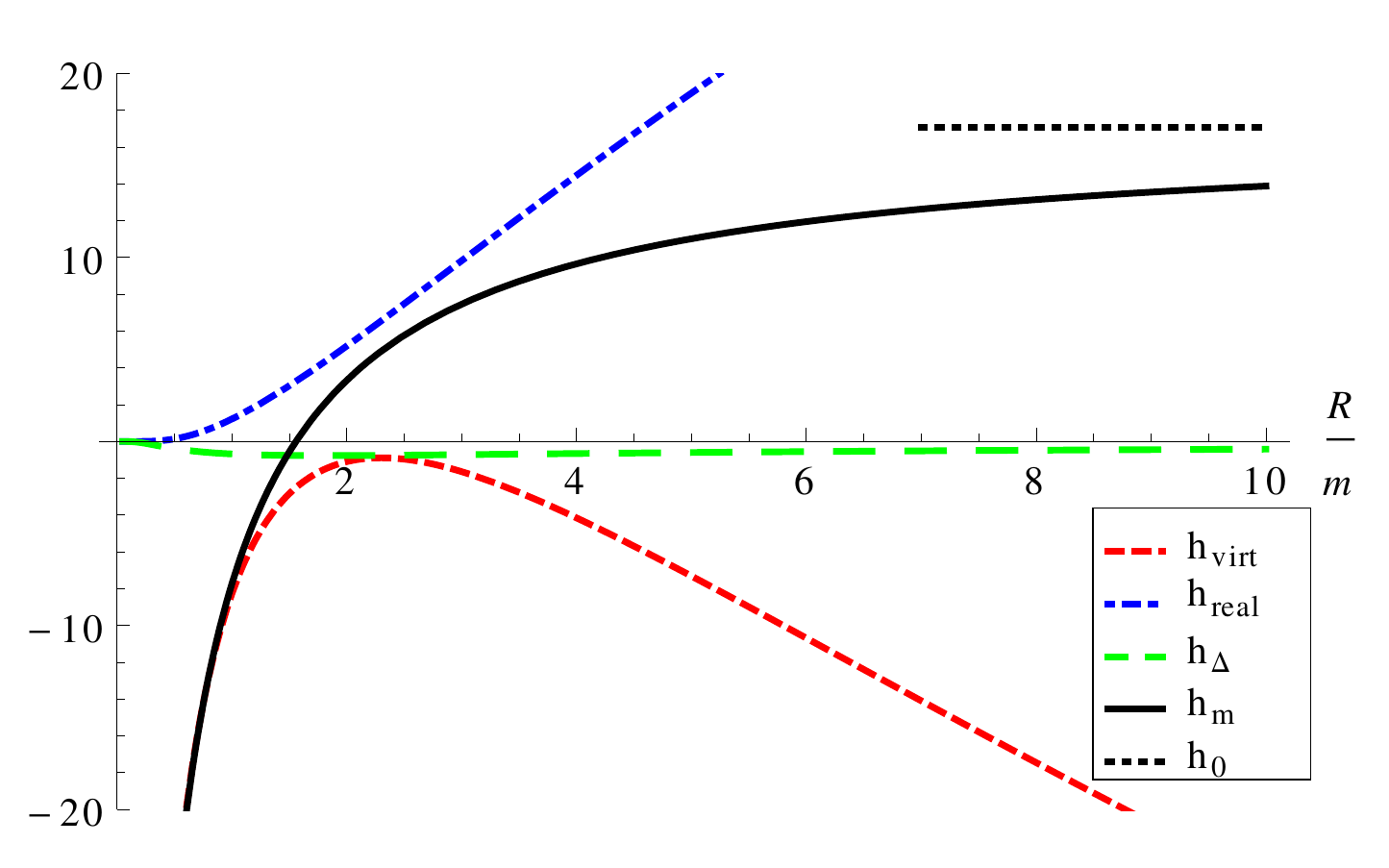} }
\caption{Gap subtractions $h_{\rm virt}(R,m,\mu=R)$, $h_{\rm real}(R,m)$ and
$h_{\Delta}(R,m)$ coming from the ${\cal O}(\alpha_s C_F T_F)$ massive quark virtual (red dashed line) and
real (blue dotted-dashed
  line) contributions to the gluon hemisphere soft function and from the
phase
  space misalignment correction (green wide-dashed line). The black solid line $h_m(R,m,\mu=R)$
denotes
  the sum of all terms and the black dotted line $h_0(R,\mu=R)$ the well known massless
limit.
  The virtual contributions fully determine the $m\to\infty$ limit.}
\label{fig:gapsubtraction}
\end{figure}
The results for $h_{\rm virt}(R,m,\mu=R)$, $h_{\rm real}(R,m)$,
$h_{\Delta}(R,m)$ and their sum are shown in Fig.~\ref{fig:gapsubtraction}
as a function of
$m/R$ and $R/m$. Note that the phase space misalignment correction $h_{\Delta}(R,m) \sim \mathcal{O}(R^6/m^6)$ for $R \ll m$ and $h_{\Delta}(R,m) \sim \mathcal{O}(m/R)$ for $R \gg m$. We see that 
$h_\Delta$, which contains only the
phase space contribution where the quark and antiquark enter different
hemispheres, is very small. This is not unexpected since
this phase space configuration is related to larger gluon invariant masses
and
therefore less sensitive to infrared renormalon-type contributions than the phase space contributions in $h_{\rm virt}$ and $h_{\rm real}$.
We also see that in the massless limit $R/m\gg 1$ there are large
cancellations between
the virtual and real radiation contributions in $h_{\rm
  virt}$ and $h_{\rm real}$. This is related to the fact that for the massive
quark corrections to the soft function real and virtual contributions each
contain mass-singularities, and the sum of both is needed
to reach the known massless limit (indicated by the black dotted line), 
\begin{align}\label{eq:deltamassless}
& h_{\rm virt}(R,m,\mu)+h_{\rm real}(R,m)\stackrel{m\rightarrow
0}{\longrightarrow} h_0(R,\mu) \nn \\
& \equiv  \frac{8}{3}
   \ln^2\left(\frac{\mu^2}{R^2}\right)  + \frac{80}{9}\ln\left(
     \frac{\mu^2}{R^2}\right) +\frac{8}{9}\pi^2 + \frac{224}{27}   \, .
\end{align}
This result agrees with Ref.~\cite{Hoang:2008fs}. Including the first
non-vanishing correction to the massless limit we obtain for the whole
renormalon subtraction\footnote{The small mass expansion yields $-8\pi z$ for $h_{\rm real}$ and $\approx -7.88 z$ for $h_\Delta$ at linear order in $z$.}
\begin{align}
 \delta_m(R,m,\mu) \stackrel{m\rightarrow 0}{\longrightarrow}
& \, \frac{\alpha_s^2
   C_F T_F}{(4\pi)^2}Re^{\gamma_E} \nn \\
&   \times \left\{h_0(R,\mu)-37.1
 \frac{m}{R}+\mathcal{O}\left(\frac{m^2}{R^2}\right)\right\}  \, .
\end{align}
The fact that a term linear in $m/R$ arises is directly tied to the
existence of the ${\cal
  O}(\Lambda_{\rm QCD})$ renormalon in the soft function indicating a linear
sensitivity to small momenta \cite{Beneke:1994bc,Hoang:1999us,Hoang:2000fm}.
For large masses the virtual
contribution gives the leading order behavior, while the real radiation contribution decouples
$h_{\rm real}(z\gg 1)=  16 \sqrt{2\pi} z^{-\frac{5}{2}} e^{-z}[1
  -\frac{49}{8z}+\mathcal{O}\left(\frac{1}{z^2}\right)]$, which leads to ($L_m=\ln(m^2/\mu^2)$)
\begin{align}\label{eq:deltamasslarge}
& \delta_m(R,m,\mu) \stackrel{m\rightarrow \infty}{\longrightarrow} 
\frac{\alpha_s^2
   C_F T_F}{(4\pi)^2}Re^{\gamma_E} \left[ h_{\rm virt}(R,m,\mu)+{\cal
O}\left(\frac{R^6}{m^6}\right)
\right] \nn \\ 
&=\frac{\alpha_s^2
   C_F T_F}{(4\pi)^2}Re^{\gamma_E}
\left[-\frac{16}{3}L_m\ln\left(\frac{\mu^2}{R^2}\right)-\frac{8}{3}
L_m^2-\frac{80}{9}L_m \right. \nn \\
& -\left. \frac{224}{27} +\mathcal{O}\left(\frac{R^6}{m^6}\right)\right]
\,.
\end{align}
So the gap subtraction does not decouple by itself, indicating that the evolution in
$R$ of
the renormalon-free gap parameter $\bar{\Delta}$ (which gives the
scale-dependence of the
nonperturbative matrix element $\Omega_1$ in the leading power corrections of
the OPE for $\ell\gg\Lambda_{\rm QCD}$ \cite{Hoang:2007vb}) has a decoupling relation when the
evolution crosses the mass threshold. This decoupling takes place
simultaneously when the massive quark is decoupled from the soft function as well as the strong coupling,
see
Refs.~\cite{Gritschacher:2013pha,Gritschacher:2013next}. 

Since $h_{\rm real}$ is cumbersome to evaluate in a numerical code and $h_\Delta$ is not known analytically, we provide a parametrization for $\delta_m(R,m,\mu)$. 
We can write
\begin{align}\label{eq:renormalon_para}
 \delta_m(R,m,\mu)= & \, \frac{\alpha_s^2 C_F T_F}{(4\pi)^2}  R e^{\gamma_E} \left[\frac{8}{3}\ln^2\left(\frac{\mu^2}{R^2}\right)+\frac{80}{9}\ln\left(\frac{\mu^2}{R^2}\right)\right] \nn \\
 &+\delta_m(R/m) \, ,
\end{align}
where $\delta_m(R/m)=\delta_m(R,m,R)$. A good parametrization for $\delta_m(R/m)$ is provided by
\begin{align}
 & \delta_m(x=R/m) =  \frac{\alpha_s^2 C_F T_F}{(4\pi)^2} R e^{\gamma_E} \left\{\left[-\frac{8}{3}
\ln^2 x^2+\frac{80}{9}\ln\, x^2\right.\right. \nn\\
&\left.-\left.\frac{224}{27}\right]\left(1-e^{-\alpha/x^{\beta}}\right)+\left[\frac{224}{27}+\frac{8}{9}\pi^2\right]e^{-\gamma/x^{\delta}}\right\} \, ,
\end{align}
which implements already the correct asymptotic behavior for small and large $x$. The four free parameters are fixed by a $\chi^2$-fit giving
\begin{align}
 & \alpha = 3.01 , \,\beta = 1.64 , \, \gamma = 4.62, \, \delta = 1.66 \, ,
\end{align}
 which approximates the exact result to better than $1\%$ for arbitrary ratios $R/m$.

The ${\cal O}(\alpha_s^2 C_F T_F)$ massive quark corrections to the gap
subtractions also give contributions to the evolution in $R$ of the subtracted
gap
parameter $\bar\Delta$, which is
free of the ${\cal O}(\Lambda_{\rm QCD})$ renormalon. Recalling that the
subtracted gap parameter is related to the
unsubtracted (and scale-independent) gap parameter $\Delta$ by the relation
\begin{align}
\Delta = \bar\Delta(R,m,\mu)+\delta(R,m,\mu)\,,
\end{align}
the R-evolution equation for the gap parameter $\bar{\Delta}$ for $\mu=R$
can be
written as
\begin{align}
R\frac{d}{d R} \bar{\Delta}(R,m,R)& =-R \frac{d}{d R}
\delta(R,m,R) \nn \\
&=-R\sum_{n=0}^{\infty} \gamma_n^R
\left(\frac{\alpha_s(R)}{4\pi}\right)^{n+1} \, .
\end{align}
The terms in the R-evolution equation up to ${\cal O}(\alpha_s^2)$ for gluonic and
massless quark corrections were determined in Ref.~\cite{Hoang:2008fs}.
The ${\cal O}(\alpha_s^2 C_F T_F)$ massive quark contributions can be
determined
from Eq.~(\ref{eq:deltam}) giving
\begin{align}
\gamma_{1,m}^R=\gamma_{1,m}^{\rm virt}+\gamma_{1,m}^{\rm
real}+\gamma_{1,m}^{\Delta}\,,
\end{align}
where
\begin{align}
&\gamma_{1,m}^{\rm  virt}=C_F T_F e^{\gamma_E}\left\{-\frac{8}{3}
\ln^2\left(\frac{m^2}{R^2}\right)
+\frac{16}{9}\ln{\left(\frac{m^2}{R^2}\right)}+\frac{256}{27}\right\} \, ,
\\
&\gamma_{1,m}^{\rm real}=C_F T_F
e^{\gamma_E}\left\{\frac{16}{3}G^{3,0}_{1,3}\left(\begin{matrix} 1 \\0,0,0
\end{matrix}\middle\vert
\frac{z^2}{4}\right)+\frac{32}{9}K_0(z) \right. \nn \\
&-\frac{32}{27}zK_1(z)+\frac{32}{27}z^2K_0(z)+\frac{16\pi}{27}
z^3 \nn\\
& \left. -\frac{16\pi}{27} z^4\left[K_1(z)L_{-2}(z)+
K_2(z)L_{-1}(z)\right]\right\} \label{eq:Revo_real}\, .
\end{align}
The term $\gamma_1^{\Delta}$ cannot be given in analytic form and has to be computed
numerically from Eq.~(\ref{eq:DeltaS_thrust_ren}). Its contribution is,
however,
very small and might be insignificant for practical applications. Alternatively, the $\mathcal{O}(\alpha_s^2 C_FT_F)$ massive quark contributions to the R-evolution can be determined from the parametrization in Eq.~(\ref{eq:renormalon_para}).

\section{Conclusions}\label{sec:conclusions}

In this paper we have completed the computation of the partonic soft function for the double hemisphere mass and thrust distribution in SCET at $\mathcal{O}(\alpha_s^2)$ by providing the $\mathcal{O}(\alpha_s^2 C_F T_F)$ corrections coming from secondary massive quarks. This has been achieved by first considering a modified phase space such that dispersion techniques could be applied allowing for a simple analytic computation. Afterwards, the UV-finite phase space misalignment corrections have been computed with numerical methods. Based on the results in Ref.~\cite{Kelley:2011ng} we have been able to use our massive quark results, which provide a well controlled regularization of IR divergences, to determine explicit results for the massless quark $\mathcal{O}(\alpha_s^2 C_F T_F n_f)$ and the gluonic $\mathcal{O}(\alpha_s^2 C_A C_F)$ corrections to the momentum space double hemisphere mass soft function in terms of distributions. These expressions have not yet been given in previous literature. Finally, to 
remove the sensitivity on infrared scales we have calculated the renormalon subtractions for the massive quark contributions in the gap 
formalism for the soft function and provided the corresponding terms in the R-evolution equation above the mass scale.  

The results in this paper are an integral part of a N$^3$LL order description of $e^+ e^-$ event shape distributions related to hemisphere masses (and thrust) which account for massive quark effects.

\appendix

\section{Computation of the cumulant for the phase space misalignment correction for thrust}\label{app:a}

\begin{figure*}
  \begin{center}
  \subfigure[$L<m$]{\epsfig{file=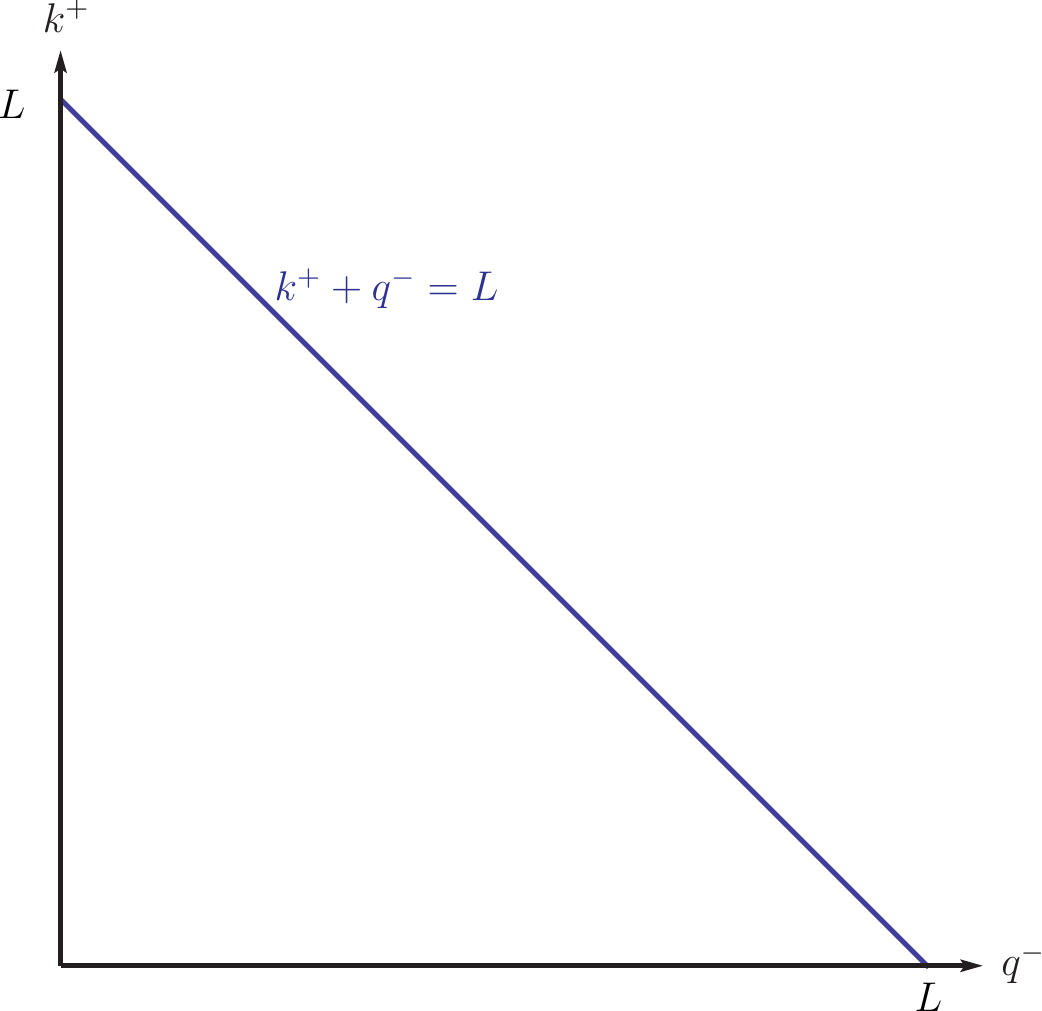,width=0.35\linewidth,clip=}}
  \subfigure[$m<L<(1+\sqrt{5})m/2$]{\epsfig{file=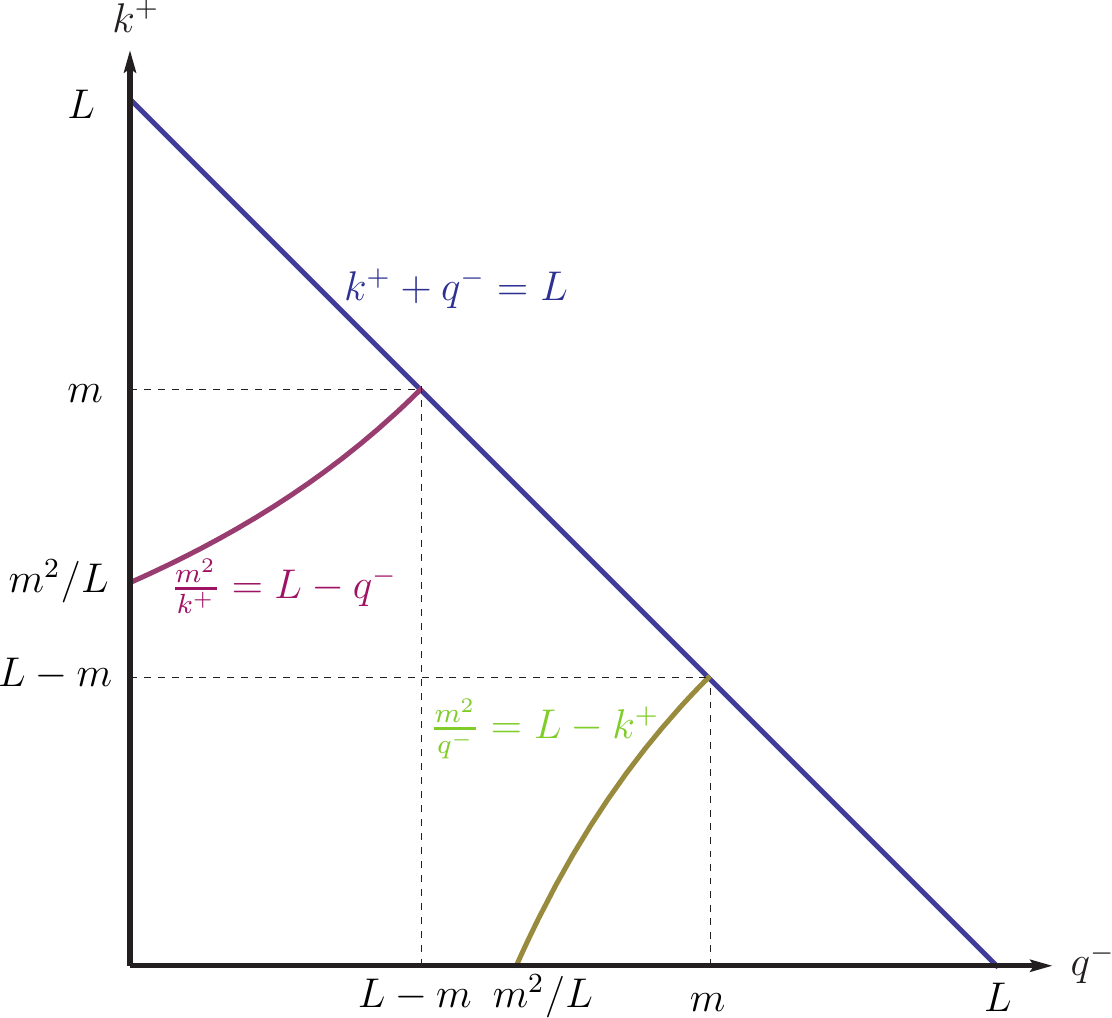,width=0.35\linewidth,clip=}}
  \subfigure[$(1+\sqrt{5})m/2<L<2m$]{\epsfig{file=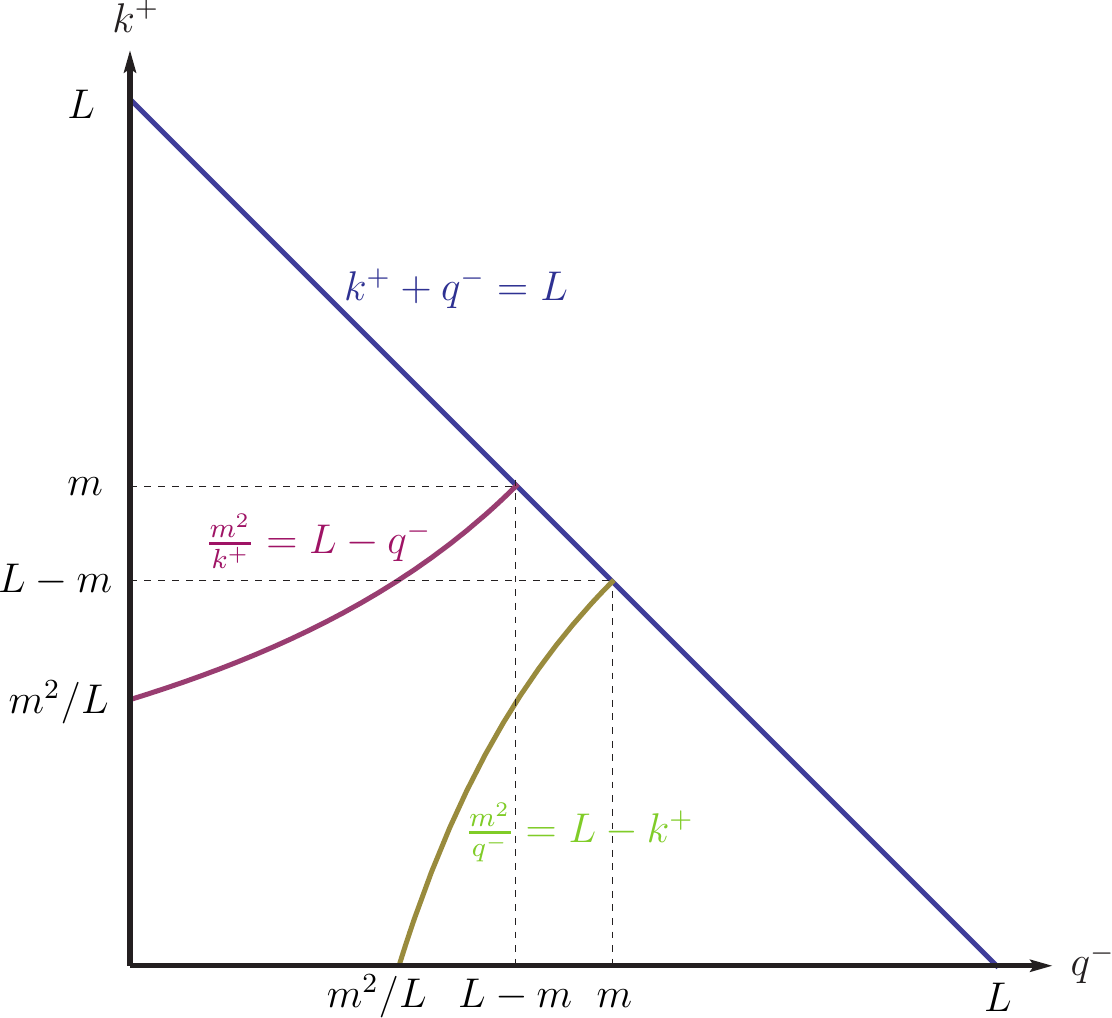,width=0.35\linewidth,clip=}}
  \subfigure[$2m<L$]{\epsfig{file=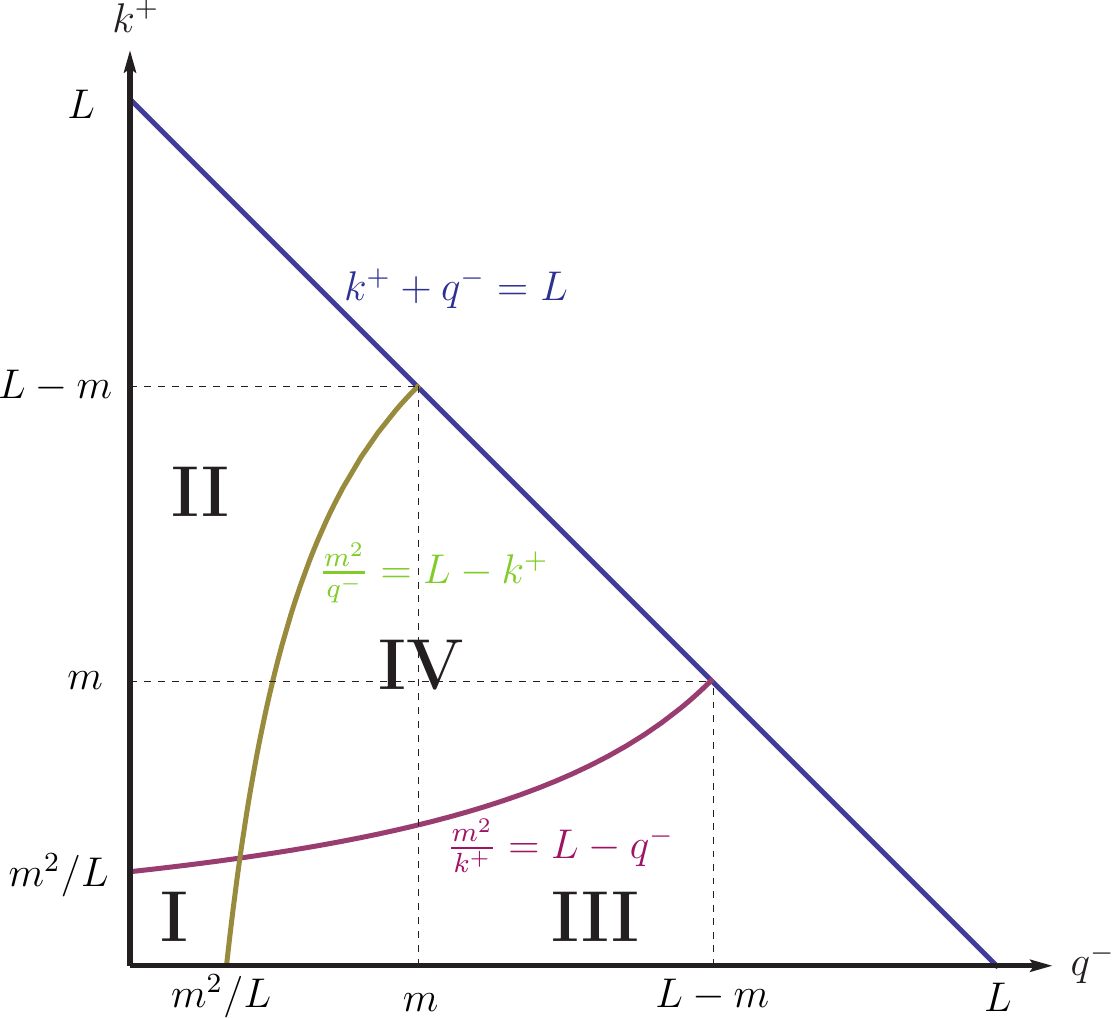,width=0.35\linewidth,clip=}}
  \caption{The integration areas for different parameter regimes. For $L<m$ we have one single integration domain. For $m<L<(1+\sqrt{5})m/2$ and $(1+\sqrt{5})m/2<L<2m$ there are three domains, the regimes differ by the hierarchy between $L-m$ and $m^2/L$. Finally, for $L>2m$ we have 4 areas, where the central one (IV) becomes dominating for large values of $L$.}
  \label{fig:Integration_regimes}
  \end{center}
\end{figure*}

Here we give some details on a computation of the phase space misalignment correction $\Delta S_{\tau}(\ell,m)$ which does not rely on the separate determination of the contributions from the quark and gluon hemisphere prescriptions in the phase space region where the two quarks enter opposite hemispheres.
We consider the cumulant 
\begin{equation}
 \Delta \mathcal{S}_{\tau} (L,m)= \int_0^L d\ell \, \Delta S_{\tau}(\ell,m) \, ,
\end{equation}
which can be rearranged onto a single integration domain using the relation
\begin{align}\label{eq:phase-space-diff}
 & \int_0^L d\ell \,\theta(k^{-}-k^{+}) \, \theta(q^{+}-q^{-})\, \left[\delta(\ell-k^{+}-q^{-}) \right. \nn\\
 & \hspace{0.5cm}-\theta(q^{+}+k^{+}-q^{-}-k^{-}) \, \delta(\ell-k^{-}-q^{-}) \nn \\
 & \hspace{0.5cm}- \left. \theta(q^{-}+k^{-}-q^{+}-k^{+}) \, \delta(\ell-k^{+}-q^{+})\right] \nn \\
 & = \theta(k^{-}-k^{+}) \, \theta(q^{+}-q^{-}) \, \theta(k^{+}+q^{+}-L) \, \theta(k^{-}+q^{-}-L) \nn \\
 & \hspace{0.5cm} \times \theta(L-k^{+}-q^{-})  \, .
\end{align}
After integration over the transverse momenta the cumulant adopts the form
\begin{align}\label{eq:DeltaS_PS}
 &\Delta \mathcal{S}_\tau (L,m)= \frac{\alpha_s^2 C_F T_F}{16\pi^2} \int\limits_0^L dq^{-} \int\limits_0^{L-q^{-}} dk^{+} \int\limits_{L-k^{+}}^{\infty} dq^{+} \int\limits_{L-q^{-}}^{\infty} dk^{-} \nn \\
 & \times \theta(k^{+}k^{-}-m^2) \, \theta(q^{+}q^{-}-m^2) f_m(k^{+},k^{-},q^{+},q^{-},m) 
\end{align}
with $f_m(k,q,m)$ given in Eq.~(\ref{eq:fm}). Note that in the massless case the on-shell constraint $\theta$-functions can be dropped and the integrations yield directly a constant corresponding to Eq.~(\ref{eq:DeltaS_massless}). In order to unambiguously determine the integration domains in the 4-dimensional integral it is convenient to distinguish between 4 parameter regimes: $L<m$, $m<L<(1+\sqrt{5})m/2$, $(1+\sqrt{5})m/2<L<2m$ and $2m<L$. We illustrate the areas with different integration domains for the larger momenta, $q^{+}$ and $k^{-}$, for each regime in Figs.~\ref{fig:Integration_regimes}a-d in the plane of the smaller momenta $q^{-}$, $k^{+}$. One integration can be performed analytically, the remaining ones can be done numerically using the Cuba library \cite{Hahn:2004fe}. Using both deterministic as well as Monte-Carlo algorithms we obtain the same result. Differentiating with respect to $L$ yields $\Delta S_\tau (\ell,m)$.

We give a short outline for the calculation of the asymptotic expansion of $\Delta S_{\tau}(\ell,m)$ for large thrust momenta.  
The asymptotic expansion is performed for each area in Fig.~\ref{fig:Integration_regimes}d with cutoff regularization taking $m^2/L \ll \Lambda_1 \ll m \ll \Lambda_2 \ll L$. Area I is suppressed by $m^6/L^6$ and thus irrelevant. For the computation of the areas II and III we obtain
\begin{align}
 &\Delta \mathcal{S}_\tau^{(II)} (L,m) = \Delta \mathcal{S}_\tau^{(III)} (L,m) =\frac{\alpha_s^2 C_F T_F}{16\pi^2} \left\{\frac{m^2}{L^2}\right. \nn \\
 &\left.\times\left[2\,\ln^2\left(\frac{m^2}{L^2}\right)+12 \, \ln \left(\frac{m^2}{L^2}\right) +12+\frac{8\pi^2}{3}\right]+ \mathcal{O}\left(\frac{m^3}{L^3}\right) \right\} \, ,
\end{align}
The expansions in area IV are cumbersome. We have to go to NLO and consider a large number of different scaling regions with difficult integrations. Special care has to be taken of the power counting for the computation with several cutoffs separating the different regions. Furthermore, cancellations in the denominator appear at NLO which require a special treatment for the hemisphere border region with $q^{-}\approx q^{+}$ and $k^{+} \approx k^{-}$. The calculation for area IV eventually yields
\begin{align}
 &\Delta \mathcal{S}_\tau^{(IV)} (L,m) = \frac{\alpha_s^2 C_F T_F}{16\pi^2} \left\{-\frac{64}{9}+\frac{104\pi^2}{27}-\frac{64\zeta(3)}{3} \right. \nn\\
 &-  \frac{m^2}{L^2}\left[8 \, \ln^2\left( \frac{m^2}{L^2}\right)+56 \, \ln \left(\frac{m^2}{L^2}\right) +\frac{296}{3}+\frac{386 \pi^2}{45} \right. \nn\\
 &\left.+\left.16\pi\right]+\mathcal{O}\left(\frac{m^3}{L^3}\right) \right\} \, .
\end{align}
Thus, the final result for the integrated soft function difference reads
\begin{align}
 & \Delta \mathcal{S}_\tau (L,m) = \frac{\alpha_s^2 C_F T_F}{16\pi^2} \left\{ -\frac{64}{9}+\frac{104\pi^2}{27}-\frac{64\zeta(3)}{3} \right. \nn \\
 &-\frac{m^2}{L^2}\left[4\,\ln^2 \left(\frac{m^2}{L^2}\right) +32 \, \ln \left(\frac{m^2}{L^2}\right) +\frac{224}{3}+\frac{146 \pi^2}{45} \right. \nn \\
 & \left.+ \left. 16\pi\right] +\mathcal{O}\left(\frac{m^3}{L^3}\right) \right\} \, ,
\end{align}
which gives Eq.~(\ref{eq:DeltaS_lowmass}).

\section{Integral representations of special functions}\label{app:b}

Here we give explicit integral representations for the Meijer G function $G^{3,0}_{1,3}$ and Struve functions $L_n$ ($n<-1/2$) used by {\it Mathematica} and appearing in Eqs.~(\ref{eq:delta_real}),~(\ref{eq:Revo_real}). For $z>0$ they read
\begin{align}\label{eq:G3013}
 &G^{3,0}_{1,3}\left(\begin{matrix} 1
\\0,0,0 \end{matrix}\middle\vert\frac{z^2}{4}\right) = 4\int_1^{\infty} \frac{dt}{t} \, K_0(z t)\, ,\\
 & L_n = I_{-n}(z)-\frac{2^{1-n} z^n}{\sqrt{\pi } \,\Gamma \left(n+\frac{1}{2}\right)} \int_0^{\infty }dt \left(t^2+1\right)^{n-\frac{1}{2}} \sin (t z) \, ,
\end{align}
where $K_n$ and $I_{n}$ indicate the better known Bessel functions. Computing the integral in Eq.~(\ref{eq:G3013}) numerically is faster and more stable than evaluating $G^{3,0}_{1,3}$ directly in {\it Mathematica} (in particular for large vales of the argument). See also \cite{Wolfram} for further information about these functions.

\bibliography{softfct}{}
\bibliographystyle{my_bibstyle}
 
\end{document}